\documentclass[12pt]{article}

\advance\voffset by -2.0cm
\advance\hoffset by -1.25cm
\usepackage{pstricks,psfrag}
\usepackage{amsmath}
\usepackage{amssymb}
\usepackage{amsthm}
\usepackage{amsfonts}

\numberwithin{equation}{section}

\theoremstyle{definition}

\newcommand{\CC}{\mathbb{C}} 
\newcommand{\RR}{\mathbb{R}} 
\newcommand{\ZZ}{\mathbb{Z}} 
\newcommand{\NN}{\mathbb{N}} 

\DeclareMathOperator{\im}{Im} 
\DeclareMathOperator{\Tr}{Tr} 
\DeclareMathOperator{\rank}{rank} 
\DeclareMathOperator{\Sp}{Sp} 

\newcommand{\J}{\mathcal{J}} 
\newcommand{\CFT}{\mathcal{H}}
\newcommand{\M}{{\mathcal M}} 
\newcommand{\Sieg}{\mathfrak{H}} 
\newcommand{\Sch}{\mathfrak{S}} 
\newcommand{\be}{\begin{equation}}
\newcommand{\ee}{\end{equation}}

\newcommand{\g}{\mathfrak{g}}

\newlength{\oldcolsep}

\title{{\LARGE Higher genus partition functions of meromorphic conformal field theories}
\vspace*{0.5cm}}

\author{
{\Large Matthias R.\ Gaberdiel}\thanks{\tt E-mail: gaberdiel@itp.phys.ethz.ch} \  and
{\Large Roberto Volpato}\thanks{\tt E-mail: volpato@itp.phys.ethz.ch}
\\ \\
{Institut f\"ur Theoretische Physik,
ETH Z\"urich} \vspace*{0.1cm} \\
{8093 Z\"urich, Switzerland} \vspace{0.3cm} \\
}
\date{\today}
\begin{document}
\maketitle

\begin{abstract}
It is shown that the higher genus vacuum amplitudes of a meromorphic
conformal field theory determine the affine symmetry of the theory
uniquely, and we give arguments that suggest that also the
representation content with respect to this affine symmetry is
specified, up to automorphisms of the finite Lie algebra. We
illustrate our findings with the self-dual theories at $c=16$ and
$c=24$; in particular, we give an elementary argument that shows
that the vacuum amplitudes of the $E_8\times E_8$ theory and the
$Spin(32)/\mathbb{Z}_2$ theory differ at genus $g=5$. The fact that
the discrepancy only arises at rather high genus is a consequence of
the modular properties of higher genus amplitudes at small central
charges. In fact, we show that for $c\leq 24$ the genus one
partition function specifies  already the partition functions up to
$g\leq 4$ uniquely. Finally we explain how our results generalise to
non-meromorphic conformal field theories.
\end{abstract}

\newpage
\renewcommand{\theequation}{\arabic{section}.\arabic{equation}}

\section{Introduction}

The genus one partition function of a conformal field theory determines
the spectrum of the theory uniquely, but there are different
conformal field theories that have the same genus one partition function.
Probably the best known example is the case of the $E_8\times E_8$ and the
$Spin(32)/\mathbb{Z}_2$ theories at $c=16$ that have the same torus
vacuum amplitude (and hence the same number of states at each
conformal weight), but that are evidently different conformal field theories (since
they have different Lie symmetries and thus have different correlation
functions).

In the context of string theory, for example in the framework of the
AdS$_3$/CFT$_2$ correspondence
\cite{Witten:2007kt,Maloney:2007ud} (see also \cite{Gaiotto:2007xh,Yin:2007gv, Yin:2007at})
one often does not have direct access to the correlation functions of the
(dual) conformal field theory that would specify the conformal field theory completely.
Instead one has control over the vacuum amplitudes at arbitrary genus.
It is then a natural question to ask to which
extent this information specifies the (dual) conformal field theory uniquely.

In this paper we shall study this question for the case of
meromorphic conformal field theories (that are relevant in the
context of \cite{Witten:2007kt,Maloney:2007ud}); the restriction to
meromorphic theories simplifies our arguments, but is not crucial
for our analysis, and essentially all our arguments  work equally
well in the general case. As we shall see, the higher genus vacuum
amplitudes always determine the Lie symmetry of the theory
completely, and we shall give arguments that suggest that the same
is true for the representation content (up to automorphisms of the
Lie algebra). As a special case, we give an elementary argument to
show  that the $E_8\times E_8$ and the $Spin(32)/\mathbb{Z}_2$
theories have different genus $g=5$ amplitudes, in agreement with
the recent analysis of \cite{Grushevsky:2008zp}.
\medskip

The basic strategy of our analysis is as follows. There is a degeneration limit of
a genus $g$ surface in which it becomes a torus with $g-1$ nodes:

\vspace{0.4cm}
\begin{figure}[htb]\begin{center}
\resizebox{\textwidth}{!}{\input{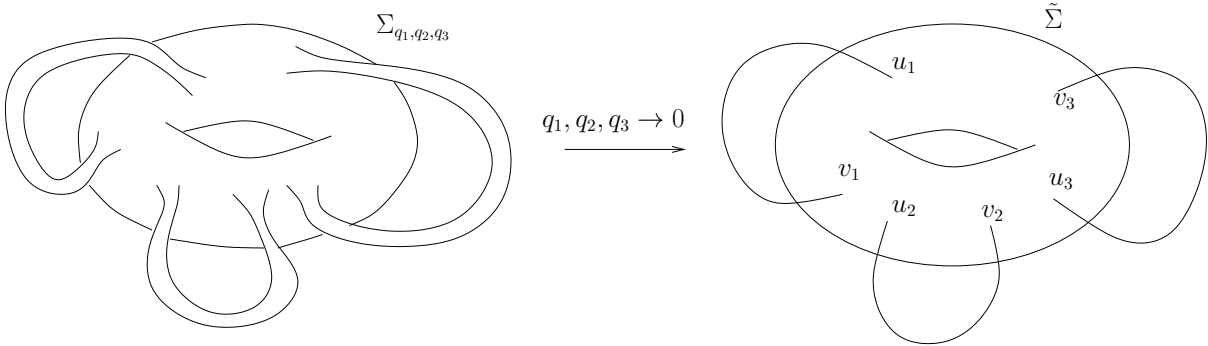}}\end{center}
\caption{\small A family of Riemann surfaces $\Sigma_{q_1,q_2,q_3}$ of genus
$4$ degenerates to a singular surface with three nodes.}
\label{f:multipledeg}\end{figure}
In this limit, the genus $g$ vacuum amplitude is described by sums of
$2(g-1)$ point functions on the torus, where we sum over an orthonormal basis
of states inserted at $(v_i,u_i)$, $i=1,\ldots g-1$, and weight the contribution
of the state with conformal dimension $h_i$ at $(v_i,u_i)$ with $q_i^{h_i}$.
If we consider the term that is proportional to $\prod q_i$, we get
a sum over $2(g-1)$ point functions of currents (fields of conformal weight one).
By integrating these
currents along one of the cycles of the torus, we can convert them into
zero modes. Thus, starting from a genus $g$ vacuum amplitude, we can
determine the trace over the full space of states, where we insert in addition to
$q^{L_0}$ also suitable combinations of  generators of the finite dimensional
Lie algebra. In fact, these combinations always define Casimir operators of the
Lie algebra, and we can determine their
eigenvalues (on the states of a given conformal dimension) from these
considerations. This allows us to determine the underlying Lie algebra completely.
We also argue, by considering more general degeneration limits,
that we can determine the representation content of the theory (up to the
ambiguity of the action of outer automorphisms) at arbitrary conformal weight.

We illustrate our findings with a number of explicit examples. In particular, we study
the self-dual conformal field theories at $c=16$ and $c=24$
\cite{Dolan:1989vr,Schellekens:1992db}, and show that
all pairs of theories that have the same genus one partition function can be
distinguished by their genus $g=5$ amplitudes. We also show that for $c=32$
such pairs of theories can typically already be distinguished at genus $g=2$, and
we give an explanation of these phenomena by studying the constraints from
modularity systematically. Among other things, this allows us to show that
for $c\leq 24$ the genus $g$ amplitudes with $g\leq 4$ are already uniquely
determined in terms of the genus $g=1$ amplitude, while no such constraint exists
at $c\geq 32$.
\medskip

Higher genus (vacuum) amplitudes of conformal field theories have been studied before
among others in
\cite{Friedan:1986ua,Bernard:1988yv,Segal:2002ei,Zhu:1994wq,Gawedzki:1994rf,%
Tuite:1999id,Mason:2007ph}. There is also some extended literature
on higher genus amplitudes in string theory, see for example
\cite{Belavin:1986ga,AlvarezGaume:1986es,Verlinde:1986kw} and the
reviews \cite{D'Hoker:1988ta,D'Hoker:2002gw,Albeverio:1997}; some more
recent progress is also described in
\cite{Cacciatori:2008ay,Grushevsky:2008zm,Grushevsky:2008qp,%
Matone:2008td,Grushevsky:2008zp}.
\bigskip

The paper is organised as follows. In section~2 we outline the general structure
of our argument. To illustrate the basic ideas we consider, in section~3,
the examples of the self-dual meromorphic fields theories at $c=16$, $24$
and $c=32$. In particular, we demonstrate  that all pairs of inequivalent theories
can be distinguished by (higher) genus amplitudes. In section~4 we analyse
the modular properties of the higher genus amplitudes systematically, and
thus explain our findings of section~3 from this perspective. In section~5
we work out the general argument that shows that higher genus amplitudes
determine the Lie symmetry uniquely. We also argue, using similar techniques,
that the same can be said about the representation content with respect to
the affine algebras. (This result relies on a Lie algebraic conjecture for which we
give some evidence in appendix~C.)
Finally, section~6 contains our conclusions where we indicate
among other things how our arguments generalise to
non-meromorphic conformal field theories. Appendix~A gives some details
of our calculations for $c=24$, while appendix~B collects some general facts
about Riemann surfaces and their Schottky covers.

\section{Partition functions and Lie algebra invariants}

Let us begin by reviewing some standard material concerning genus
$g$ partition functions.

\subsection{Partition functions and degeneration
limits}\label{s:partfunct}

In this paper we shall consider self-dual meromorphic conformal
field theories, {\it i.e.}\ theories that are purely left-moving.
These theories arise, for example, as the left-moving part of a
holomorphically factorising conformal field theory, as in
\cite{Witten:2007kt}. As we have mentioned before (see also the
conclusions), our arguments also work for more general conformal
field theories, but the restriction to meromorphic conformal field
theories will simplify some of our notation considerably.

We shall always assume that the theory has a unique vacuum state $\Omega$ of conformal
dimension zero, and that the spectrum of $L_0$ is a subset of the non-negative integers.
This allows us to write
\be
\CFT = \bigoplus_{h=0}^{\infty} \CFT_h \ , \ee where $\CFT_h$ is the
subspace of states of $L_0$ eigenvalue $h$. We shall always assume
that each eigenspace $\CFT_h$ is finite dimensional. The genus one
partition function of the theory then equals the genus one character
$\chi_{g=1}(\tau)$, which is a holomorphic function of the modulus
$\tau$ of the torus. The usual modular consistency condition
requires that $\chi_{g=1}(\tau)$ is modular invariant; if we think
of the meromorphic conformal field theory to be the left-moving part
of a holomorphically factorising theory, the character only has to
be modular invariant up to a phase. In either case, the modular
$S$-matrix is essentially trivial, and hence Verlinde's formula
implies that the meromorphic conformal field theory has only one
representation, namely $\CFT$ itself.

The genus $g$ analogue of the chiral character $\chi_{g=1}(\tau)$
defines a holomorphic section $\chi_g$ in the line bundle
$\lambda^{c/2}$ that is defined on the moduli space $\M_g$ of
Riemann surfaces of genus $g$. Here $\lambda$ is the determinant
line bundle and $c$ the central charge (see, for example,
\cite{Segal:2002ei} for details). Again, for the left-moving part of
a holomorphically factorising theory, the genus $g$ partition function
$\chi_g$ must satisfy appropriate modular properties under
$Sp(2g,{\mathbb Z})$; this will be described in more detail in section~4.

\begin{figure}[htb]\begin{center}
\resizebox{\textwidth}{!}{\input{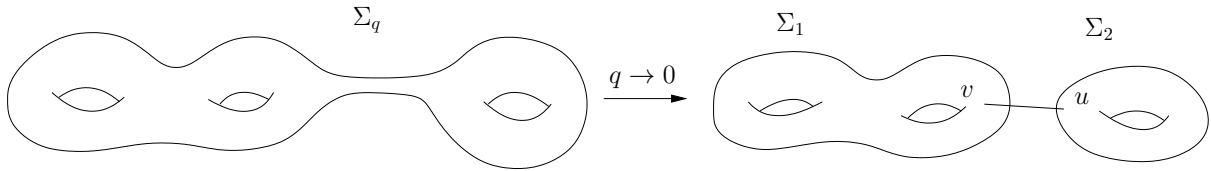}}\end{center}
\caption{\small By a separating degeneration limit of a family of
smooth Riemann surfaces, a singular Riemann surface with  node (here
represented by a line) is obtained. The surface is given by two
smooth components $\Sigma_1$ and $\Sigma_2$ of genus $k$ and $g-k$
with marked points $u\in\Sigma_1$ and $v\in\Sigma_2$ joined by a
node.}\label{f:sepdeg}\end{figure}

\begin{figure}[htb]\begin{center}
\resizebox{\textwidth}{!}{\input{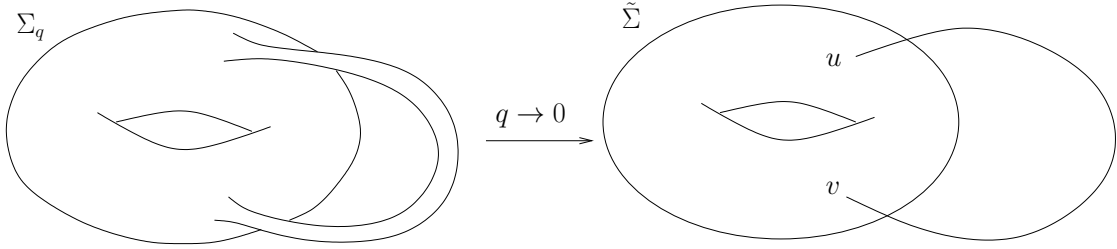}}\end{center}
\caption{\small A singular Riemann surface obtained by a
non-separating degeneration limit. The points $u,v$ on a surface
$\tilde\Sigma$ of genus $g-1$ are identified to form a node (here
represented by a line).}\label{f:nonsepdeg}\end{figure}

The genus $g$ partition function $\chi_g$ also satisfies certain
factorisation relations. Let $\Sigma_q$ be a family of Riemann
surfaces that degenerate in the limit $q\rightarrow 0$. There are
two cases of interest: first, a homologically trivial cycle can be
pinched down to a node. In this case the limit $q\rightarrow 0$
describes a union of two connected components, $\Sigma_1$ and
$\Sigma_2$ of genus $k$ and $g-k$, $1\le k\le [g/2]$, respectively
(see figure~\ref{f:sepdeg}).
 The other case is that a
homologically non-trivial cycle is pinched down, in which case the
degenerate limit surface has genus $g-1$ (see
figure~\ref{f:nonsepdeg}).
 In either
case, the genus $g$ partition function converges to the partition
function of the appropriate limiting surface. For example, in the
second case where a homologically non-trivial cycle is pinched, the
partition function becomes
\be\label{lead}
\chi_g(\Sigma_q) \stackrel{q\to 0}{\longrightarrow}
\chi_{g-1}(\tilde\Sigma)\ , \ee where $\tilde\Sigma$ is the surface
obtained from the singular curve by removing the node. The overall
normalisation of the partition functions is fixed by $\chi_{g=0}=1$.

Equation (\ref{lead}) describes the leading behaviour as
$q\rightarrow 0$, but one can also be more specific about the
subleading terms. In fact, in any such degeneration limit, the
chiral partition function $\chi_g$ can be expanded in a power series
in the degeneration parameter $q$ (see \cite{Friedan:1986ua})
$$
\chi_g = \sum_{h=0}^{\infty} q^{h}\sum_{i\in I_h} \left\langle
V(\psi^{(h)}_i,u) \, V(\psi^{(h)}_i,v) \right\rangle_{\tilde\Sigma}\
,
$$
where $h$ labels the eigenvalues of the $L_0$ operator (conformal
weights) in $\CFT$,  and the $\psi^{(h)}_i$, ${i\in I_h}$ are an
orthonormal basis for the states $\CFT_h$ of conformal weight
$h$.\footnote{The power series on the right hand side converges for
sufficiently small $q$.} Furthermore, $V(\phi,z)$ denotes the vertex
operator corresponding to the state $\phi$, and $u,v\in\tilde\Sigma$
are the points on the (possibly disconnected) Riemann surface
$\tilde\Sigma$ that are identified by the node to form the singular
surface $\Sigma_0$.

In the following we shall be interested in the particular case of
multiple degenerations in which a Riemann surface of genus $g$
becomes a surface of genus $1$ with $g-1$ nodes (see figure~1 in the
introduction). In this case it is useful to regard the partition
function as a holomorphic function on the Schottky space
\cite{Bernard:1988yv}
 (see appendix~\ref{a:schottky} for more
details). The degeneration limit we are considering corresponds to
the limit in which $g-1$ out of $g$ multipliers
$q_{1},\ldots,q_{g-1}$ of the Schottky group generators vanish, so
that, upon setting $q\equiv q_g$, we obtain
\begin{equation}\label{e:expand}
\chi_g = \sum_{h_1,\ldots,h_{g-1}} \prod_{j=1}^{g-1}
q_j^{h_j}\sum_{i_1,\ldots,i_{g-1}} \Tr\left( \prod_{j={1}}^{g-1}
V(\psi^{(h_j)}_{i_j},u_j)\, V(\psi^{(h_j)}_{i_j},v_j)\, q^{L_0}
\right)\ .
\end{equation}
Note that the standard definition of the genus $1$ character as the trace of
the operator $q^{L_0-c/24}$ is related to $\chi_1(q)$
as
\be \label{e:chi1}
\chi_1(q)=q^{c/24}\Tr(q^{L_0-{c/24}})\  . \ee The extra factor
$q^{c/24}$ is due to the conformal transformation (see
\eqref{e:flattoSchottky}) from the cylinder to the annulus. With
this definition, $\chi_1$ is smooth in the limit $q\to
0$, which corresponds to the degeneration of a torus to a sphere.

\subsection{Lie algebra considerations}

In the following we shall mainly be interested in the
contribution to (\ref{e:expand}) from states at $h_j=1$. We therefore
need to review what is known about these states in general.

In any meromorphic conformal field theory, the states at conformal
weight $h=1$ give rise to an affine Lie algebra symmetry (see for example
\cite{Goddard} for a more detailed exposition). Indeed if we denote the fields
of conformal dimension one (the currents) by $J^a(z)$, then their operator product
expansion is necessarily of the form
\be\label{OPE}
J^a(z)\, J^b(w) =
\frac{\kappa^{ab}}{(z-w)^2}+f^{ab}_{\phantom{ab}c}\frac{J^c(w)}{z-w} +{\cal O}(1) \ ,
\ee
where $\kappa^{ab}$ and $f^{ab}_{\phantom{ab}c}$ are constants. Defining
the modes of these fields via
\be\label{modes}
J^a_n=\oint dz\, z^{n}J^a(z) \ ,
\ee
it follows from (\ref{OPE}) that they satisfy the commutation relations of
an affine Kac-Moody algebra $\hat\g$
\be
[J^a_m,J^b_n]=m\, \kappa^{ab}\, \delta_{m,-n}+f^{ab}_{\phantom{ab}c}\,
J^c_{m+n}\ .
\ee
Note that the zero modes $J^a_0\equiv t^a$ form a finite-dimensional Lie algebra $\g$
whose structure constants are given by $f^{ab}_{\phantom{ab}c}$. Furthermore,
$\kappa^{ab}$ is a symmetric tensor that is invariant with respect to $\g$. If the
conformal field theory is unitary, then $\kappa^{ab}$ is positive definite, and thus the
finite dimensional Lie algebra $\g$ is semi-simple, or a direct sum of simple Lie algebras
and some $u(1)$ factors.

In each simple factor, $\kappa$ is proportional to the Cartan-Killing form $K^{ab}$ of the
Lie algebra $\g$. We choose the standard convention for the normalisation of the
Cartan-Killing form, namely that the longest roots of the Lie algebra
have length squared equal to $2$. Furthermore, we pick a basis for the
Lie generators of $\g$ such that $K^{ab}=\delta^{ab}$. With these conventions
$\kappa^{ab}$ is then of the form
\be\label{Lienorm}
\kappa^{ab} = k\, \delta^{ab} \ ,
\ee
where $k$ is the level that takes a specific fixed value for each simple factor.
If we assume that the theory is unitary then each $k$ must be a positive integer.
The coefficient of the identity in the OPE (\ref{OPE}) determines the normalisation
of the currents; at level $k$, the currents $J^a$ have therefore norm $k$. In order
to have an orthonormal basis we therefore have to rescale them as
\be\label{Jres}
\hat{J}^a = k^{-\frac{1}{2}} \, J^a \ .
\ee
In the following the quadratic Casimir operator of the finite dimensional Lie
algebra will play an important role. We choose the (usual) convention that
the quadratic Casimir $C_2$ is given by
\be
C_2 = \sum_a t^a t^a \ .
\ee
In the adjoint representation the value of $C_2$ is then equal to
$2 h^\vee(\g)$, where  $h^\vee(\g)$ is the dual Coxeter number of
the finite dimensional Lie algebra $\g$, and for the simply-laced
algebras we have
\be
h^\vee (a(n)) = n+1 \ , \quad h^\vee (d(n)) = 2n-2 \ , \quad h^\vee
(e6) = 12 \ , \quad h^\vee(e7) = 18 \ , \quad h^\vee(e8) = 30 \ .
\ee For the rescaled $\hat{J}^a$ generators it then follows that
\be
\label{e:Liealgnorm} \Tr_{ad}(\hat{J}^a_0
\hat{J}^a_0)=\frac{2h^\vee(\g)}{k}\, \dim(\g) \ .
\end{equation}

\subsection{Lie algebra invariants in degeneration limits}

After this interlude we are ready to return to the degeneration
limits of genus $g$ partition functions. Let us consider the coefficients
of (\ref{e:expand}) that contain at most linear powers of $q_i$, {\it i.e.}
\begin{align}\label{e:Zexpansion}\chi_g
=& \Tr\left( q^{L_0} \right) +
\sum_{i=1}^{g-1}q_i\sum_a\Tr\left(\hat{J}^a(u_i)\hat{J}^a(v_i)\, q^{L_0}\right)\\
&+\sum_{i\neq j}q_iq_j\sum_{a,b}
\Tr\left(\hat{J}^{a}(u_i)\, \hat{J}^{a}(v_i)\, \hat{J}^{b}(u_j)\, \hat{J}^{b}(v_j)\, q^{L_0}\right)\notag\\
&+\ldots+q_1\ldots q_{g-1}
\sum_{a_1,\ldots,a_{g-1}}\Tr\left(\prod_{i=1}^{g-1}\hat{J}^{a_i}(u_i)\,
\hat{J}^{a_i}(v_i)\, q^{L_0}\right) +{\cal O}(q_i^2)\notag\ ,
\end{align}
where ${\cal O}(q_i^2)$ is a term of order $2$ in at least one of
the parameters $q_1,\ldots,q_{g-1}$. The functions that appear on
the right hand side are correlation functions of currents
\be\label{2.16}
\sum_{a_1,\ldots,a_l}\Tr\left(\prod_{i=1}^l
\hat{J}^{a_i}(u_i)\hat{J}^{a_i}(v_i)q^{L_0}\right)\ .
\ee
If we know the vacuum amplitude at genus $g$, we can thus determine all
these correlation functions, where the number of currents, $2l$, is less or equal
than $2(g-1)$ (and the modular parameter of the torus $\tau$ is arbitrary). These
amplitudes depend obviously on the Lie group symmetries of
the theory, as well as its representations content. The simplest way to make
this dependence explicit is to integrate the insertion points
$u_1,v_1,\ldots,u_l,v_l$ along the $\alpha$-cycle of the torus. Because of
(\ref{modes}) this then replaces the current $\hat{J}^a$ by its zero mode,
$\hat{J}^a_0$. In doing these integrals, there is a choice corresponding to the
ordering of the integrals. Thus we may take the $2l$ zero modes to appear in any
order. The simplest ordering is the one where the two
zero modes $\hat{J}^{a_i}_0$ stand next to each other, {\it i.e.}\ the term of the form
\be\label{e:integrampl}
\sum_{a_1,\ldots,a_l}\Tr\left(\hat{J}_0^{a_1}\,
\hat{J}_0^{a_1}\cdots \hat{J}_0^{a_l}\, \hat{J}_0^{a_l}\, q^{L_0}
\right) = \frac{1}{k^l}\, \Tr\left(C_2^l \, q^{L_0} \right) \ . \ee
Since $\hat{J}^a_0$ commutes with $L_0$, the coefficient of $q^n$ in
this series comes from the states of conformal weight $n$, $\CFT_n$.
Let us decompose $\CFT_n$ in terms of irreducible representations of
$\g$ as
\be
\CFT_n = \bigoplus_{R} m_{n,R} \, R \ ,
\ee
where $m_{n,R}$ is the multiplicity with which the irreducible representation
$R$ appears in $\CFT_n$. If we denote the value of the quadratic Casimir $C_2$ in
$R$ by $C_2(R)$, then we can rewrite (\ref{e:integrampl}) as\footnote{For simplicity
of notation we are assuming here that all simple factors of $\g$ have the same
level $k$; otherwise we need to rescale the currents of the different simple factors
differently.}
\be\label{2.19}
\sum_{a_1,\ldots,a_l}\Tr\left(\hat{J}_0^{a_1}\,
\hat{J}_0^{a_1}\cdots \hat{J}_0^{a_l}\, \hat{J}_0^{a_l}\, q^{L_0}
\right) = \sum_n q^n \, \sum_R m_{n,R} \, \frac{C_2(R)^l}{k^l}
\dim(R) \ . \ee The genus $g$ partition function thus determines
these generating series for any $l\leq g-1$. More generally, by
choosing a different ordering for the integrals, the genus $g$
partition function also determines the expressions
\be\label{2.20}
\sum_{a_1,\ldots,a_{2l}}\Tr\left(\hat{J}_0^{a_{\sigma(1)}}\,
\hat{J}_0^{a_{\sigma(2)}}\cdots \hat{J}_0^{a_{\sigma(2l-1)}}\,
\hat{J}_0^{a_{\sigma(2l)}}\, q^{L_0}
\right)\prod_{i=1}^l\delta_{a_ia_{i+l}} \ , \ee where $\sigma$ is
any permutation in $S_{2l}$ (and again $l\leq g-1$). In analogy to
(\ref{2.19}) the coefficient of $q^n$ in (\ref{2.20}) can then be
expressed in terms of (in general higher order) Casimir operators.
\smallskip

In the following we shall study the information that can be obtained
in this manner systematically. In particular, we shall show
(see section~\ref{s:geninvariants}) that these amplitudes determine
the affine Lie algebra that is defined by the currents uniquely.
Before we delve into this analysis, it may be instructive to study
a few simple cases first.

\section{Applications and results}\label{s:results}

It follows from the considerations of the previous section that the
genus $g$ vacuum amplitude determines the expression (\ref{2.19}). In particular,
if the genus $g$ partition function of two meromorphic conformal field theories
agrees, so must the expressions (\ref{2.19}) for $l\leq g-1$. For many theories
the right hand side of (\ref{2.19}) can be evaluated  fairly easily. Thus we may
turn the logic around: if (\ref{2.19}) is different for two conformal field theories
for a given $l$, then the genus $g=l+1$ vacuum amplitude of the two theories must
be different. In this section we shall apply these ideas to meromorphic conformal
field theories at $c\leq 32$.

In all examples we have considered we find that the theories can be distinguished
by some higher genus  vacuum amplitude. For small values of the central charge
({\it i.e.}\ for $c\leq 24$), we typically have to go up to genus $g\geq 5$ in order
to distinguish theories; for $c=32$, on the other hand, the discrepancy typically
occurs already at genus $g=2$. This behaviour is a consequence of the structure
of higher genus modular forms; this will be explained in section~\ref{s:modular}.
\smallskip

Self-dual meromorphic conformal field theories only exist at central
charges that are integer multiples of $8$ \cite{Goddard}. The simplest examples are the
theories of $c$ chiral bosons on an even unimodular lattice $\Lambda$ of
rank $c$. For such theories, the sub-lattice $\Lambda_2\subseteq \Lambda$
generated by its elements
of length squared two is the root lattice of some Lie algebra $\g$, and the
theory corresponding to $\Lambda$ contains the affine Kac-Moody algebra $\hat\g$ at
level $1$ as a subalgebra. In most cases the theories therefore have an interesting
Lie symmetry, and the constraints coming from (\ref{2.19}) are powerful.

For $c=8$ and $c=16$, it is believed that all self-dual conformal field theories
are such lattice theories. In fact, for $c=8$, the only self-dual conformal field theory is
believed to be the lattice theory based on the $e8$ root lattice $\Gamma_{e8}$;
this theory is equivalent to the $e8$ level $k=1$ affine vertex operator algebra (VOA).
The situation is more interesting for $c=16$ where two self-dual theories are known
(and believed to be the only self-dual theories): the lattice theory based on
$\Gamma_{e8} \oplus \Gamma_{e8}$ that is equivalent to the
$e8\oplus e8$ affine VOA  at level one and that is often referred to as the
$E_8\times E_8$ theory. And the lattice theory based on
$\Gamma_{16}$, whose sublattice $\Lambda_2$ is the root lattice of $so(32)$.
The latter VOA contains the $\g=so(32)$ affine VOA at level $k=1$ as a
proper subalgebra. At  conformal weight $2$ this VOA contains a chiral spinor
representation of $so(32)$, and thus the Lie group symmetry is $Spin(32)/\ZZ_2$
(rather than $SO(32)$).

For $c\geq 24$, on the other hand, there are additional self-dual conformal field
theories that can be obtained as a $\ZZ_2$ orbifold from the lattice
theories, see in particular \cite{Dolan:1989vr} for explicit constructions at $c=24$.
However, even at  $c=24$, it is not believed that these lattice and orbifold theories
already account for all self-dual conformal field theories. In fact
Schellekens \cite{Schellekens:1992db} has conjectured that there are additional
self-dual conformal field theories whose genus $g=1$ partition function and
Lie symmetry he determined. The situation for $c\geq 32$ is less clear; there is
already a gigantic number of lattice theories, and they probably only describe a small
subset of all the self-dual theories.

In the following we shall study the behaviour of the higher genus amplitudes
for the theories at different values of the central charge in turn.

\subsection{The two self-dual theories at $c=16$}

As mentioned before, at $c=16$ there are two different self-dual conformal field
theories, the $E_8\times E_8$ theory based on $\Gamma_{e8} \oplus \Gamma_{e8}$,
and the $Spin(32)/\ZZ_2$ theory based on $\Gamma_{16}$. It is well known that their
genus one amplitudes agrees; in particular, this implies that the graded dimensions
$\dim\CFT_h$ of the $E_8\times E_8$ theory and the $Spin(32)/\ZZ_2$ theory
are equal for all values of $h$. At $h=1$, the former theory contains the
$248+248$ states coming from $e_8\oplus e_8$, while the latter theory
contains the $496$ states coming from the adjoint representation of $so(32)$.
With respect to this Lie symmetry we can then decompose also the states at higher
conformal weight. For example, at $h=2$, the $E_8\times E_8$ theory contains
the states
\be\label{e8de}
E_8\times E_8: \qquad \CFT_2= \Bigl[{\bf 1}\otimes({\bf 1}\oplus
{\bf 248}\oplus {\bf 3875})\Bigr]\oplus \Bigl[({\bf 1}\oplus {\bf
248}\oplus {\bf 3875})\otimes {\bf 1}\Bigr]\oplus \Bigl[{\bf
248}\otimes {\bf 248}\Bigr] \ , \ee where we have denoted the
different $e8$ representations by their dimension; in particular,
{\bf 248} ist the adjoint representation, and the Dynkin labels of
${\bf 3875}=[1,0,0,0,0,0,0,0]$.\footnote{We are using the same
labelling for the Dynkin labels as LiE.} For later convenience we
also give the values of the quadratic Casimirs
\be
\begin{array}{ll}
C_2({\bf 1}\otimes {\bf 1})=0 \ , \qquad
& C_2({\bf 1}\otimes {\bf 248}) = C_2({\bf 248}\otimes {\bf 1}) = 60  \\
C_2({\bf 248}\otimes {\bf 248}) = 120 \qquad &
C_2 ({\bf 1}\otimes {\bf 3875}) = C_2({\bf 3875}\otimes {\bf 1}) =96 \ .
\end{array}
\ee
\smallskip

\noindent For $Spin(32)/\ZZ_2$ the decomposition is
\be\label{so32de}
Spin(32)/\ZZ_2: \qquad
\CFT_2\equiv {\bf 1}\oplus
{\bf 496}\oplus {\bf 527}\oplus {\bf 35960}\oplus {\bf 32768}\ ,
\ee
where, in terms of Dynkin labels
\begin{align*}
{\bf 1}\equiv &[0, 0, 0, 0, 0, 0, 0, 0, 0, 0, 0, 0, 0, 0, 0,0]\qquad
C_2({\bf 1}) = 0 \\
{\bf 496}\equiv&[0, 1, 0, 0, 0, 0, 0, 0, 0, 0, 0, 0, 0, 0, 0, 0]\qquad
C_2 ({\bf 496}) = 60 \\
{\bf 527}\equiv&[2, 0, 0, 0, 0, 0, 0, 0, 0, 0, 0, 0,0, 0, 0, 0]  \qquad C_2({\bf 527})=64\\
{\bf 35960}\equiv &[0, 0, 0, 1, 0, 0, 0, 0, 0, 0, 0, 0, 0, 0, 0, 0] \qquad C_2({\bf 35960})=112\\
{\bf 32768}\equiv &[0, 0, 0, 0, 0, 0, 0, 0, 0, 0, 0, 0, 0, 0, 0, 1]\qquad C_2({\bf 32768})=124 \ ,
\end{align*}
and we have again given the eigenvalues of the quadratic Casimir in each case.
One easily checks that the total dimension of $\CFT_2$ is the same
in both cases (namely $69752$).
\smallskip

It has been known for some time that the vacuum amplitudes of the
$E_8\times E_8$ and $Spin(32)/\ZZ_2$ theories are the same for $g\le 4$.
Recently, it has been proved that the two partition functions are different for
$g=5$ \cite{Grushevsky:2008zp}. We want to give an elementary argument for
this, using the techniques we have developed above. From what we have
said above, the fact that the partition functions are equal for $g\leq 4$ must in particular
mean that the trace of $C_2^l$ must agree for $l=1,2,3$. On the other hand,
if (\ref{2.19}) was different for $l=4$, this would imply that the genus $g=5$
amplitudes differ.

Let us study (\ref{2.19}) for the first few powers of $q$. At $q^1$, the states
in $\CFT_1$ contribute. Both theories have $k=1$, and thus the relevant
expressions are
\be
\begin{array}{ll}
Spin(32)/\ZZ_2: \quad &
\Tr_{\CFT_1}(C_2^l)=\Tr_{ad}(C_2^l)=\dim(so(32))\, 2^l\, h^\vee (so(32))^l \\[6pt]
E_8\times E_8: \quad &
\Tr_{\CFT_1}(C_2^l)=2\Tr_{ad}(C_2^l)=2\dim(e8)\, 2^l\, h^\vee(e8)^l
\ .
\end{array}
\ee
Since $\dim(so(32)) = 496 =  2 \dim(e8)$, and
$h^\vee (so(32)) = 30 = h^\vee(e8)$ it follows that there is no discrepancy for
any $l$.

The situation is however different at $q^2$. Given the values of the quadratic Casimir
operators given above, it is straightforward to calculate the trace of $C_2^l$
on $\CFT_2$. Explicitly,
\be
\begin{array}{ll}
Spin(32)/\ZZ_2: \quad &
\Tr_{\CFT_2}(C_2^l) = 1\cdot 0^l + 496 \cdot 60^l + 527 \cdot 64^l + 35960\cdot 112^l
+ 32768 \cdot 124^l \\[6pt]
E_8\times E_8: \quad &
\Tr_{\CFT_2}(C_2^l) = 2 \cdot 0^l +
2 \cdot 248\cdot 60^l + 2 \cdot 3875\cdot 96^l + 248\cdot 248\cdot 120^l \ .
\end{array}
\ee
One then finds that the two expressions agree for $l=1,2,3$, but disagree for
$l=4,5,\ldots$. In particular, this provides an independent (and elementary) proof
that the two partition functions disagree for $g=5,6,\ldots$. Our analysis is also
compatible with the known fact that they agree for $g\leq 4$.

\subsection{The self-dual theories at $c=24$}

There are $24$ even unimodular lattices (Niemeier lattices) of rank $24$,
each one corresponding to a distinct meromorphic conformal field theory.
The theory based on the Leech lattice, has an abelian Lie algebra symmetry
$u(1)^{24}$, whereas in all the other cases the Lie algebra is
non-abelian and semi-simple.

If two such theories have a different number of currents, the
partition function is obviously different already at genus $g=1$. On
the other hand, modular invariance of the genus $1$ character
implies (see section \ref{s:modular}) that the genus $1$ partition
function for the lattice $\Lambda$ depends only on the number
$N=N_\Lambda$ of currents, {\it i.e.}\ on the number of elements
of length squared two in the lattice $\Lambda$. Among the
$24$ Niemeier lattices, there are five pairs of lattices that have the same
number $N_\Lambda$; they are listed in table \ref{t:Niemeier} (as customary,
Niemeier lattices are denoted by the Lie algebras whose root lattice is
generated by the elements of length squared two).
\begin{table}[htb]\begin{center}\begin{tabular}{|c|c|c|c|c|c|c|c|c|}
\hline&&&&&&&&\\[-13pt] $\Lambda$ &  $d{24}$ & $d{16}\, e8$ & $(e8)^3$ &
$a{24}$ & $(d{12})^2$&
$a{17}\, e7$ & $d{10} \, (e7)^2$& $a{15}\,d9$\\
\hline $N_\Lambda$ & 1128 & \multicolumn{2}{|c|}{744} &  624 & 552 &
\multicolumn{2}{|c|}{456} & 408\\
$h^\vee_\Lambda$ & $46$ & \multicolumn{2}{|c|}{$30$} & $25$ & $22$ &
\multicolumn{2}{|c|}{$18$} &  $16$\\
\hline\hline&&&&&&&&\\[-13pt]
$\Lambda$ & $(d8)^3$  &$(a{12})^2$ & $a{11}\, d7\, e6$ & $(e6)^4$
& $(a9)^2\, d6$ & $(d6)^4$ & $(a8)^3$ &$(a7)^2\, (d5)^2$\\
\hline $N_\Lambda$ & 360 &
336 & \multicolumn{2}{|c|}{312} & \multicolumn{2}{|c|}{264}& 240  & 216 \\
$h^\vee_\Lambda$ & $14$ & $13$ & \multicolumn{2}{|c|}{$12$} &
\multicolumn{2}{|c|}{$10$} & $9$ & $8$ \\
 \hline\hline&&&&&&&&\\[-13pt] $\Lambda$ & $(a6)^4$ &
$(a5)^4d4$&$(d4)^6$ & $(a4)^6$&$(a3)^8$  &$(a2)^{12}$ &
$(a1)^{24}$ & $u(1)^{24}$\\
\hline $N_\Lambda$ & 192 & \multicolumn{2}{|c|}{168} & 144 & 120 & 96 & 72&24\\
$h^\vee_\Lambda$ & $7$ & \multicolumn{2}{|c|}{$6$}&  $5$ & $4$
& $3$ & $2$ & -\\
\hline\end{tabular}\caption{{\small Niemeier lattices $\Lambda$,
number $N_\Lambda$ of currents and dual Coxeter number
$h^\vee_\Lambda$ of each simple Lie algebra
factor.}}\label{t:Niemeier}\end{center}\end{table}

In all cases (except the Leech lattice) the Lie algebra $\g=\oplus \g_i$ is the direct
sum of simply laced simple Lie algebras $\g_i$.  Furthermore, the dual Coxeter
number is  the same for all the simple algebras that appear in a given lattice,
$h^\vee_\Lambda=h^\vee(\g_i)$ for all $i$. For any simply laced simple Lie algebra $\g$,
the dual Coxeter number $h^\vee(\g)$ is related to the rank $r(\g)$ and
the dimension $\dim(\g)$ of $\g$ as
\be
r(\g) \Bigl( h^\vee(\g)+1 \Bigr) =\dim(\g) \ .
\ee
For the Lie algebras $\g$ appearing in the Niemeier lattices, the total rank of $\g$
is always $24$, and hence
\be
N_\Lambda \equiv\dim(\g)=\sum_i\dim(\g_i)
=\sum_ir(\g_i)(h^\vee(\g_i)+1)
=(h^\vee_\Lambda+1)\sum_i r(\g_i)=24(h^\vee_\Lambda+1)\ .
\ee
Thus $h^\vee_\Lambda$  actually only depends on $N_\Lambda$  as
$h_\Lambda=\tfrac{N_\Lambda}{24}-1$, and hence
\be
\Tr_{\CFT_1}(C_2^l)=\sum_i (2h^\vee(\g_i))^l\,
\dim(\g_i)=(2h_\Lambda^\vee)^l \, N = \left( \frac{N_\Lambda}{12} -
2 \right)^l\, N_\Lambda \ , \ee so that two theories with the same
number of currents cannot be distinguished by the trace
$\Tr_{\CFT_1}(C_2^l)$, for any $l$. As in the case of the $c=16$
theories, let us therefore consider the trace of the powers of the
quadratic Casimir over $\CFT_2$. The results can be determined from
the decomposition of $\CFT_2$ in terms of representations of $\g$
(see appendix~\ref{H2deco}), and are given in
table~\ref{t:NiemTraces}.\smallskip

It is striking that in all cases $\Tr_{\CFT_2}(C_2^l)$ agrees for
$l=1,2,3$, but disagrees for $l=4$. As in the situation at $c=16$ this
proves that the partition functions are different for genus $g=5$. It also
suggests that the partition functions may be the same for $g\leq 4$. We shall
prove that this is in fact so in section~\ref{s:modular}.

\begin{table}[htb]\begin{center}\begin{tabular}{|c|c|c|c|c|}\hline&&&&\\[-13pt]
$N_\Lambda=744$ &$d{16}\, e8$ & $(e8)^3$ &
difference & g\\ \hline $\dim(\CFT_2)$ & 196884& 196884& 0 & 1\\
$\Tr_{\CFT_2}(C_2)$ &23302080& 23302080& 0 & 2\\
$\Tr_{\CFT_2}(C_2^2)$ &2766787200&  2766787200& 0 & 3\\
$\Tr_{\CFT_2}(C_2^3)$ &329282496000& 329282496000& 0 & 4\\
$\Tr_{\CFT_2}(C_2^4)$ &39259975772160&  39257415936000& 2559836160 & 5\\
\hline
%
\hline&&&&\\[-13pt] $N_\Lambda=456$&$a{17}\, e7$ & $d{10}\, (e7)^2$ &
difference & g\\ \hline
$\dim(\CFT_2)$ &196884& 196884& 0& 1\\
$\Tr_{\CFT_2}(C_2)$ &14544576& 14544576& 0& 2 \\
$\Tr_{\CFT_2}(C_2^2)$ &1077611904& 1077611904& 0& 3 \\
$\Tr_{\CFT_2}(C_2^3)$ &80016837120& 80016837120& 0 & 4\\
$\Tr_{\CFT_2}(C_2^4)$ &5952213614592& 5952029755392& 183859200 & 5\\
\hline
%
\hline&&&&\\[-13pt] $N_\Lambda=312$ &$a{11}\, d7\, e6$ & $(e6)^4$ &
difference & g\\
\hline $\dim(\CFT_2)$ &196884& 196884& 0 &1\\
$\Tr_{\CFT_2}(C_2)$ &10041408& 10041408& 0 &2\\
$\Tr_{\CFT_2}(C_2^2)$ &513437184& 513437184& 0&3\\
$\Tr_{\CFT_2}(C_2^3)$ &26303367168& 26303367168& 0&4\\
$\Tr_{\CFT_2}(C_2^4)$ &1349589196800& 1349565235200& 23961600&5\\
\hline%
%
\hline&&&&\\[-13pt] $N_\Lambda=264$ & $(a{9})^2\, d6$ & $(d6)^4$ & difference &g \\ \hline
$\dim(\CFT_2)$ & 196884 & 196884 & 0 & 1 \\
$\Tr_{\CFT_2}(C_2)$ & 8521920 & 8521920 & 0 & 2 \\
$\Tr_{\CFT_2}(C_2^2)$ & 369747840 & 369747840 & 0 & 3 \\
$\Tr_{\CFT_2}(C_2^3)$ & 16071221760 & 16071221760 & 0 & 4 \\
$\Tr_{\CFT_2}(C_2^4)$ & 699537653760 & 699528529920 &  9123840 & 5 \\
\hline
%
\hline&&&&\\[-13pt] $N_\Lambda=168$ &$(a{5})^4\, d4$ & $(d4)^6$ &
difference & g\\ \hline
$\dim(\CFT_2)$ & 196884 & 196884 & 0 & 1 \\
$\Tr_{\CFT_2}(C_2)$ & 5455296 & 5455296 & 0 & 2 \\
$\Tr_{\CFT_2}(C_2^2)$ & 151466112 & 151466112 & 0 & 3 \\
$\Tr_{\CFT_2}(C_2^3)$ & 4211633664 & 4211633664 & 0 & 4 \\
$\Tr_{\CFT_2}(C_2^4)$ & 117240496128 & 117239851008  & 645120 & 5 \\
\hline
\end{tabular}\caption{\small Traces $\Tr_{\CFT_2}(C_2^l)$ for
CFTs corresponding to Niemeier lattices ($c=24$). We compare the
results between theories with the same number of currents
$N_\Lambda$.}\label{t:NiemTraces}\end{center}\end{table}

It is interesting to apply the same analysis also to theories that are not
lattice theories, in particular, to the $\ZZ_2$ orbifold theories constructed in
\cite{Dolan:1989vr}. The orbifold theory with affine Kac Moody symmetry
$\widehat{d9}_{2} \ \widehat{a7}_1$ has the same number of currents
($N=216$)  as the lattice theory $(a7)^2 (d5)^2$, and similarly for the
orbifold theory with affine symmetry $\widehat{d8}_{2} \ (\widehat{b4}_1)^2$
and the lattice theory $(a6)^4$ ($N=192$). The explicit results for the trace
of $C_2^l$ over $\CFT_2$ are described in table~\ref{t:DGMtheories} and it
shows exactly the same behaviour as for the pairs of Niemeier lattice theories.
\smallskip

\begin{table}[hbt]\begin{center}\begin{tabular}{|c|c|c|c|c|}
\hline&&&&\\[-13pt]  $N=216$ &$a7\, (d5)^2$ & $\widehat{d9}_2 \
\widehat{a7}_1^{}$ &
difference & g\\ \hline $\dim(H_2)$ & 196884 & 196884 & 0 & 1 \\
$\Tr_{\CFT_2}(C_2)$ & 6993216 & 6993216 & 0 & 2 \\
$\Tr_{\CFT_2}(C_2^2)$ & 248949504 & 248949504 & 0 & 3 \\
$\Tr_{\CFT_2}(C_2^3)$ & 8876805120 & 8876805120 & 0 & 4 \\
$\Tr_{\CFT_2}(C_2^4)$ & 316928581632 & 316924952832 & 3628800 & 5 \\
\hline \hline&&&&\\[-13pt] $N=192$ &$(a6)^4$ & $\widehat{d8}_2 \ \widehat{b4}_1^2$ & difference & g\\
\hline
$\dim(\CFT_2)$ &196884 & 196884 & 0 & 1 \\
$\Tr_{\CFT_2}(C_2)$ & 6225408 & 6225408 & 0 & 2 \\
$\Tr_{\CFT_2}(C_2^2)$ & 197266944 & 197266944 & 0 & 3 \\
$\Tr_{\CFT_2}(C_2^3)$ & 6260610048 & 6260610048 & 0 & 4 \\
$\Tr_{\CFT_2}(C_2^4)$ & 198933288960 & 198929660160 & 3628800 & 5 \\
\hline
\end{tabular}\caption{\small Comparison between $\ZZ_2$-twisting
theories of \cite{Dolan:1989vr} and lattice theories ($c=24$) with the same number
of currents.}\label{t:DGMtheories}\end{center}\end{table}

The pattern also continues for the theories that were conjectured to exist in
\cite{Schellekens:1992db}. If we include these theories into our considerations,
then there are many more cases where the genus $g=1$ partition functions
agree. For example the theories with affine Lie symmetry
$\widehat{e7}_3\oplus \widehat{a5}_1$ and
$\widehat{e6}_2\oplus \widehat{c5}_1 \oplus \widehat{a5}_1$ have the same
number of currents ($N=168$) as the lattice theories $(a5)^4 d4$ and $(d4)^6$.
Again, we have compared the trace of $C_2^l$ in $\CFT_2$, and the results
are described in table~\ref{t:Schelltheories}.

\begin{table}\begin{center}
\begin{tabular}{|c|c|c|c|c|c|}\hline&&&&\\[-13pt] $N=168$&$\widehat{e7}_3\ \widehat{a5}_1$ &
$\widehat{e6}_2 \ \widehat{c5}_1 \ \widehat{a5}_1$ &
$(a5)^4 d4$&
$(d4)^6$ &g\\
\hline $\dim(\CFT_2)$ & 196884 & 196884 & 196884 & 196884 &  1 \\
$\Tr_{\CFT_2}(C_2)$ & 5455296 & 5455296 & 5455296 & 5455296 &  2 \\
$\Tr_{\CFT_2}(C_2^2)$ & 151466112 & 151466112 & 151466112 & 151466112 & 3 \\
$\Tr_{\CFT_2}(C_2^3)$ & 4211633664 & 4211633664 & 4211633664 & 4211633664 &  4 \\
$\Tr_{\CFT_2}(C_2^4)$ & 1172{\bf 376}28928 & 1172{\bf 394}59328 &
1172{\bf 404}96128 & 1172{\bf 398}51008 &  5 \\
\hline
\end{tabular}\caption{{\small Comparing two of the theories of \cite{Schellekens:1992db}
with lattice theories.}}\label{t:Schelltheories}\end{center}\end{table}

Summarising our findings, it appears that we can distinguish self-dual conformal
field theories with $c\leq 24$ by determining their vacuum partition function at
genus $g=5$. On the other hand, the genus $g$ partition functions with $g\leq 4$
always seem to agree if the two theories in question have the same central
charge and the same number of currents (and hence the same torus partition
function). In section~\ref{s:modular}, we shall explain this phenomenon by studying
the constraints of modular invariance and factorisation systematically. In fact, we shall
be able to show that for $c\leq 24$ the partition functions at low genera are uniquely
determined by the number of currents.

As the central charge increases, such constraints become weaker. In
particular, for $c=32$, only the genus $1$ partition function is
completely determined by the number of currents $N$. One may then expect
that the discrepancies between partition functions of different theories already
occur for lower genera. We have tested this idea by comparing the
partition functions for a few pairs of lattice theories that have the same number of
currents, and our findings suggest that for $c=32$ different theories typically
have already different genus $g=2$ partition functions (see
table~\ref{t:c32theories}).\footnote{Note however, that for the pair
$(a1)^4 (a5)^4 d8$ and $(a3)^6 (d7)^2$ the discrepancy only seems to appear
at genus $g=3$. At $c=32$ the lattices are not uniquely determined
by their Lie algebras any more; in particular, there are more than one theories whose
Lie symmetry is $(d8)^4$. The entries in table~5 are insensitive to which of these
theories one considers, but one can distinguish them using the methods
of section~5.}
At $c=32$ the simple algebras $\g_i$ that appear in
$\g=\oplus \g_i$ have different dual Coxeter numbers, and one thus
expects that it is already sufficient to compare the traces of $C_2^l$ over
$\CFT_1$ (rather than $\CFT_2$). This is indeed borne out by our analysis
(see table~\ref{t:c32theories}).

\begin{table}[htb]\begin{center}\begin{tabular}{|c|c|c|c|c|}
 \hline&&&&\\[-13pt] $N=240$ &$(a3)^4\, (d{5})^4$ & $(a3)^8\, d8$ & difference & g\\
 \hline $\dim(\CFT_1)=N$ & 240 & 240 & 0 & 1 \\
$\Tr_{\CFT_1}(C_2)$ & 3360 & 4320 & - 960 & 2 \\
$\Tr_{\CFT_1}(C_2^2)$ & 49920 & 101760 & - 51840 & 3 \\
\hline $\dim(\CFT_2)$ &  199024 & 199024 & 0 & 1 \\
$\Tr_{\CFT_2}(C_2)$ & 5735040 & 5258880 & 476160 & 2 \\
$\Tr_{\CFT_2}(C_2^2)$ & 167260800 & 149961600 & 17299200 & 3 \\
\hline
 \hline&&&&\\[-13pt] $N=272$ & $(a3)^6(d7)^2$ & $(a1)^4(a5)^4d8$ &
difference & g\\
\hline $\dim(\CFT_1)=N$ & 272 & 272 & 0 & 1 \\
$\Tr_{\CFT_1}(C_2)$ & 5088 & 5088 & 0 & 2 \\
$\Tr_{\CFT_1}(C_2^2)$ & 110592 & 114432 & - 3840 & 3 \\
\hline $\dim(\CFT_2)$ & 206960 & 206960 & 0 & 1 \\
$\Tr_{\CFT_2}(C_2)$ & 6387072 & 6387072 & 0 & 2 \\
$\Tr_{\CFT_2}(C_2^2)$ & 205022592 & 206266752 & - 1244160 & 3 \\
\hline
 \hline&&&&\\[-13pt] $N=480$ &$(d8)^4$ & $(a1)^2\, (a9)^2\, d{12}$ &
difference & g\\
\hline $\dim(\CFT_1)=N$ & 480 & 480 & 0 & 1 \\
$\Tr_{\CFT_1}(C_2)$ & 13440 & 16128 & - 2688 & 2 \\
$\Tr_{\CFT_1}(C_2^2)$ & 376320 & 613632 & - 237312 & 3 \\
\hline $\dim(\CFT_2)$ & 258544 & 258544 & 0 & 1 \\
$\Tr_{\CFT_2}(C_2)$ & 15048960 & 13715712 & 1333248 & 2 \\
$\Tr_{\CFT_2}(C_2^2)$ & 878476800 & 757969920 & 120506880 & 3 \\
 \hline
\end{tabular}\caption{{\small Comparison between some lattice conformal field theories
at $c=32$ with the same number of currents.}}\label{t:c32theories}\end{center}\end{table}

\section{Modular properties of partition functions}\label{s:modular}

In the previous section, we compared pairs of meromorphic conformal
field theories of the same central charge and with the same number of
currents. The general behavior seems to depend on the central
charge: for $c\le 24$ the partition functions first differ at genus $g=5$, whereas
for $c=32$ the difference generically already appears at genus $g=2$.
In this section we analyse the consistency conditions of the partition
functions, in particular, modular invariance and factorisation properties,
systematically.  We shall show that for self-dual theories
with $c\leq 24$ the number of currents determines the partition functions
for genera $g\leq 4$ uniquely. On the other hand, for $c=32$, the number of
currents only determines the genus $g=1$ partition function.

\subsection{Generalities}

In general, the genus $g$  partition function of a (not necessarily
meromorphic) conformal field theory is not a function on the moduli
space $\M_g$, but rather a section of the line bundle
$\lambda^{c/2}\otimes \bar \lambda^{\bar{c}/2}$, where $\lambda$ is the
determinant line bundle on $\M_g$. In particular, for a meromorphic
conformal field theory, the generalized character $\chi$ is a
holomorphic section of the holomorphic line bundle\footnote{We
observe that $\lambda^{c/2}$ is a well-defined line bundle on $\M_g$
only if $c$ is multiple of $4$, which is the case for meromorphic
conformal field theories. In the other cases, it can only be defined as a
\emph{projective} line bundle \cite{Friedan:1986ua,Segal:2002ei}.}
$\lambda^{c/2}$.

\smallskip

The determinant line bundle $\lambda$ can be described as follows.
Consider the vector bundle $\Lambda_g$ of rank $g$ on $\M_g$, whose
fiber at the point corresponding to the Riemann surface $\Sigma$ is
the $g$-dimensional vector space of holomorphic
$1$-differentials on $\Sigma$. As shown in appendix \ref{a:Riemann},
the choice of a symplectic basis for the first homology group
$H_1(\Sigma,\ZZ)$ determines a basis $\{\omega_1,\ldots,\omega_g\}$
of holomorphic $1$-differentials on $\Sigma$, and hence
a basis of local sections on $\Lambda_g$, which we also denote
by $\omega_1,\ldots,\omega_g$.
The determinant line bundle $\lambda$ is then defined as the $g$-th
exterior product of $\Lambda_g$, and given a choice of a basis for
$H_1(\Sigma,\ZZ)$,  $\omega_1\wedge\ldots\wedge\omega_g$
defines a local holomorphic section in $\lambda$.
%
Under a symplectic transformation (see appendix~B.1)
the corresponding local section of $\lambda$ transforms as
\be
\omega_1\wedge\ldots\wedge\omega_g\ \mapsto\
\det(C\Omega+D)^{-1}(\omega_1\wedge\ldots\wedge\omega_g)\ ,\quad \hbox{where}
\quad
\left(\begin{matrix} A & B\\
C&D
\end{matrix}\right)\in\Sp(2g,\ZZ) \ . \ee
The generalised character $\chi_g$ of a meromorphic CFT is a global
holomorphic section of $\lambda^{c/2}$, so that it can be written
locally as
$$\chi_g=W_g(\Omega)\, (\omega_1\wedge\ldots\wedge\omega_g)^{c/2}\ ,
$$
where $W_g$ is a holomorphic function on the space $\J_g\subset
\Sieg_g$ of period matrices of Riemann surfaces. Since the
section cannot depend on the choice of the local trivialization,
$W_g$ must transform as a modular form of weight $c/2$
\be\label{modutW}W_g\Bigl((A\Omega+B)(C\Omega+D)^{-1}\Bigr) =\det(C\Omega+D)^{c/2}
\, W_g (\Omega)\ ,\ee under the action of $\left(\begin{smallmatrix}
A & B\\ C&D
\end{smallmatrix}\right)\in\Sp(2g,\ZZ)$. In particular, for lattice
theories, the function $W_g$ is given by
\be
W_g^{\Lambda}(\Omega)=\Theta_{\Lambda}^{(g)}(\Omega)\ ,
\ee
where
\be
\Theta_{\Lambda}^{(g)}(\Omega)=
\sum_{\lambda_1,\ldots, \lambda_g\in\Lambda} e^{\pi i \sum_{i,j}^g
\Omega_{ij}(\lambda_i,\lambda_j)} \ee is the degree $g$ theta series
of $\Lambda$.
\medskip

In section \ref{s:partfunct}, we considered the generalised
character as a holomorphic function on the Schottky space $\Sch_g$.
As explained in appendix \ref{a:schottky}, the space $\Sch_g$ of
normalised Schottky groups is a finite covering $\Sch_g\to\M_g$ of
the moduli space. The choice of a Schottky group uniformising the
Riemann surface $\Sigma$ canonically determines a set of
$\alpha$-cycles and hence a basis $\omega_1,\ldots,\omega_g$ on
$\Sigma$. This implies that the pull-back of the determinant line
bundle $\lambda_g$ to $\Sch_g$ is isomorphic to the trivial line
bundle. Thus, the only ambiguity in the identification of $\chi_g$
with a holomorphic function on the Schottky space amounts to the
choice of  a trivialisation. For our purposes we only need the $g=1$
result
\be\label{e:chi1WF}
\chi_1= q^{\frac{c}{24}} \,
(\eta^2)^{-c/2} W_1\ , \ee where
\be\eta(\tau)=q^{\frac{1}{24}}\prod_{m=1}^\infty (1-q^m)\ ,\qquad
q=e^{2\pi i\tau}
\ee
is the Dedekind eta-function. For example, for
the conformal field theory corresponding to the unimodular lattice $\Lambda$,
this formula reproduces the known result
\be
\chi^\Lambda_1=q^{\frac{c}{24}}\, \eta^{-c}(\tau)\,
\Theta_\Lambda^{(g=1)}(\tau)\ . \ee
\medskip

Apart from these modular properties, the partition function $W_g(\Omega)$
must also obey factorisation constraints.
Let us consider a family $\Sigma_t$ of Riemann
surfaces of genus $g$ that, in the limit $t\to 0$, degenerate to a
singular surface given by two components of genus $k$ and $g-k$
joined by a node. At leading order in the degeneration parameter,
the local section $(\omega_1\wedge\ldots\wedge\omega_g)^{c/2}$
factorises
\be
(\omega_1\wedge\ldots\wedge\omega_g)^{c/2}
\to(\omega_1\wedge\ldots\wedge\omega_k)^{c/2}
\otimes(\omega_{k+1}\wedge\ldots\wedge\omega_{g})^{c/2}\ ,
\ee
where
$\omega_1,\ldots,\omega_k$ and $\omega_{k+1},\ldots,\omega_{g}$ are
holomorphic $1$-differentials on the components of genus $k$ and
$g-k$, respectively. The Riemann period matrix of such a
singular surface is simply block-diagonal
\be\label{degen}
\lim_{t\to 0}\Omega_t=\Omega_{k,g-k}\equiv
\begin{pmatrix}\Omega^{(k)} & 0\\ 0 & \Omega^{(g-k)} \end{pmatrix}\ ,
\ee where $\Omega^{(k)}$ and $\Omega^{(g-k)}$ are the period
matrices of the two components. The matrix $\Omega_{k,g-k}$
corresponds to an element of the boundary of the compactification
$\bar \J_g$ in $\Sieg_g$. This implies that, in the limit
$\Omega\to\Omega_{k,g-k}$, taken along any path in $\J_g\subseteq
\Sieg_g$, $W_g$ factorises as
\be\label{Wfact}
\lim_{t\to 0}W_g(\Omega_t)= W_{g-k}(\Omega^{(g-k)})\,
W_k(\Omega^{(k)})\ . \ee Finally, since the vacuum is unique, we
have the normalisation condition
\be\label{Wnorm}
\lim_{\tau\rightarrow i\infty} W_1(\tau) = 1 \ . \ee
\smallskip

Before we analyse these constraints in more detail, it is useful to
introduce some notation. For a general modular form $f_g$ of degree
$g$ we can always consider the degeneration limit (\ref{degen}); in
this limit we can always write
\be
\lim_{t\to 0}f_g(\Omega_t)=  f_g\begin{pmatrix}\Omega^{(k)} & 0\\ 0
& \Omega^{(g-k)}
\end{pmatrix}= f_k(\Omega^{(k)})\, f_{g-k}(\Omega^{(g-k)})\ ,
\ee where $f_k$ and $f_{g-k}$ are modular forms of degree $k$ and
$g-k$, respectively. We shall use the symbolic notation
\be
f_g\to f_k\otimes f_{g-k} \ee for this factorisation property. It is
also useful to introduce the Siegel operator $\Phi$, mapping modular
forms of degree $g$ to modular forms of degree $g-1$; it is defined
by
\be
(\Phi (f_g))(\Omega^{(g-1)})=\lim_{\tau\to i\infty}f_g\begin{pmatrix}\tau & 0\\
0 & \Omega^{(g-1)} \end{pmatrix}\ . \ee The operator $\Phi$ is
linear and is compatible with the product of modular forms
\be
\Phi(f_g \,h_g)=\Phi(f_g)\, \Phi(h_g)\ . \ee The elements of its
kernel, {\it i.e.}\ the modular forms $f_g$ such that $\Phi(f_g)=0$
are called cusp forms of degree $g$. Note that if a modular form
$f_g$ of degree $g$ factorises as $f_g\to f_1\otimes f_{g-1}$ in the
limit $\Omega\to \Omega_{1,g-1}$, then
\begin{equation}\label{e:Siegfact}
\Phi(f_g)=\Phi(f_1)\, f_{g-1}\ .
\end{equation}
In particular, using (\ref{Wfact}) and (\ref{Wnorm}), it follows
that
\be\label{Wconst}
\Phi (W_g)=W_{g-1} \ee for each $g\ge 1$.

\subsection{The case of low genera $g\leq 3$}

Let us first concentrate on the case where the genus $g$ satisfies
$g\leq 3$. (We shall come back to the case of $g=4$ below.) In this
case the closure of the locus of Riemann period matrices
$\bar{\J}_g$ coincides with the Siegel upper half space $\Sieg_g$, and
thus $W_g$ must be a Siegel modular form (see appendix~\ref{a:Riemann}).
The theory of Siegel modular forms is well developed for $g\leq 3$, and we can thus be
fairly  explicit. Let us first review the salient features that will
be important for us.
\medskip

\noindent {\bf Genus $g=1$:} At genus $g=1$, the ring of modular forms is generated by the
Eisenstein series
\be
\phi_4\ , \quad \phi_6\ ,
\ee
of weight $4$ and $6$, respectively. We choose the convention that the leading term of
both $\phi_4$ and $\phi_6$ is $1$, {\it i.e.}\ that
\be
\Phi(\phi_4) = \Phi(\phi_6) = 1 \ .
\ee
Then the discriminant of the elliptic curve
\be
\Delta = \frac{\phi_4^3 - \phi_6^2}{1728} =
\eta^{24}=q-24q^2+252q^3-1472q^4+\ldots \ ,\qquad q=e^{2\pi i \tau}
\ee is a cusp form of weight $12$ (since its leading coefficient
vanishes). In fact, $\Delta$ generates the ideal of cusp forms at
genus $g=1$.
\medskip

\noindent {\bf Genus $g=2$:} The ring of modular forms of degree $g=2$ is
generated by \cite{Igusa:1962}
\be
\psi_4\ , \quad \psi_6\ ,\quad \chi_{10}\ ,\quad \chi_{12} \ .
\ee
In our conventions, the Siegel operator satisfies
\be
\Phi(\psi_4) = \phi_4 \ , \qquad \Phi(\psi_6) = \phi_6 \ , \qquad
\Phi(\chi_{10}) = \Phi(\chi_{12}) = 0 \ ,
\ee
and thus $\chi_{10}$ and $\chi_{12}$ are cusp forms. Furthermore, we have the
factorisation properties
\be
\psi_4\to \phi_4\otimes\phi_4\ , \qquad \psi_6\to
\phi_6\otimes\phi_6\ , \qquad \chi_{10}\to 0\ ,\qquad \chi_{12}\to
\Delta\otimes\Delta\ . \ee It is useful to define the modular form
$\psi_{12}=(\psi_4^3-\psi_6^2)/1728$ of weight $12$, which satisfies
the properties \be \Phi(\psi_{12})=\Delta\ ,\qquad \psi_{12}\to
\phi_4^3\otimes\Delta+\Delta\otimes\phi_4^3-1728\,\Delta\otimes\Delta\
, \ee as follows from a simple computation.
\medskip

\noindent {\bf Genus $g=3$:} The ring of modular forms is generated by $34$ modular
forms;  the generators with weight not greater than $12$ are \cite{Tsuyumine:1986}
\be
\alpha_4\ ,\quad \alpha_6\ ,\quad \alpha_{10}\ ,\quad
\alpha_{12}\ , \quad \beta_{12}\ .
\ee
We choose the conventions that the Siegel operator acts as
\be
\Phi(\alpha_4) = \psi_4 \ , \qquad \Phi(\alpha_6) = \psi_6 \ ,
\qquad \Phi(\alpha_{10}) = \chi_{10} \ , \qquad \Phi(\alpha_{12}) =
\chi_{12} \ , \qquad \Phi(\beta_{12}) = 0 \ , \ee and hence
$\beta_{12}$ is a cusp form. Furthermore, in the limit where the
genus $g=3$ surface degenerates into two surfaces of $g=2$ and
$g=1$, we have the factorisation properties
\be
\alpha_4\to \psi_4\otimes\phi_4\ , \quad \alpha_6\to
\psi_6\otimes\phi_6\ , \quad \alpha_{10}\to\chi_{10}\otimes
\phi_4\phi_6\ , \quad
\alpha_{12}\to\chi_{12}\otimes\phi_4^3+\psi_{12}\otimes\Delta , \ee
as well as
\be
\beta_{12}\to\chi_{12}\otimes\Delta\ .
\ee We also define the modular form
$\tilde\alpha_{12}=(\alpha_4^3-\alpha_6^2)/1728$, which satisfies
the properties
\be\Phi(\tilde\alpha_{12})=\psi_{12}\ ,\qquad
\tilde\alpha_{12}\to\psi_{12}\otimes\phi_4^3+\psi_4^3\otimes\Delta
-1728\,\psi_{12}\otimes\Delta\ . \ee
\bigskip

We have now collected all the relevant material to discuss the
constraints on $W_g$ that come from (\ref{modutW}) together with its
factorisation property (\ref{Wfact}). The analysis depends on the
value of the central charge, so we need to do the analysis for the
different cases separately.

\subsubsection{The case $c=8$ and $c=16$}

For $c=8$, $W_g$ is a modular form of weight $4$, while for $c=16$ the modular weight
of $W_g$ is $8$. For $g\leq 3$ there is always a unique modular form of weight four
and eight, respectively, and hence $W_g$ must be proportional to that modular form.
Using the constraint (\ref{Wconst}) as well as (\ref{Wnorm}) it then follows that
\be\label{c816}
W_1=\phi_4^{c/8}\ ,\qquad
W_2=\psi_4^{c/8}\ ,\qquad
W_3=\alpha_4^{c/8}\ .
\ee
Since for $c=8$ one such theory is the theory based on the $e8$ lattice,
it follows that we must have the identifications
\be
\phi_4=\Theta_{e8}^{(g=1)}\ ,\qquad \psi_4=\Theta_{e8}^{(g=2)}\
,\qquad \alpha_4=\Theta_{e8}^{(g=3)}\ , \ee where $\Theta_{e8}$ is
the theta series for the $e8$ lattice. In fact, by \eqref{e:chi1WF},
we can compute the partition function of the $E_8$ theory,
$\chi_1^{e8}$, using this approach, and we reobtain the known result
\be
\chi_1^{e8}=\frac{q^{1/3}}{\Delta^{1/3}}\phi_4
=q^{1/3}j(\tau)^{1/3}=1+248q+4124q^2+\ldots\ , \ee where
\be
j(\tau)=\frac{\phi_4^3}{\Delta}=\frac{1}{q}+744+196884q+\ldots \ee
is the $j$-invariant.
\smallskip

For $c=16$, on the other hand, there are two self-dual theories, namely the
$E_8\times E_8$ and the $Spin(32)/\ZZ_2$ theories. The above argument
implies that both must have the same partition function for $g=1,2,3$, namely
the one given by (\ref{c816}). This obviously ties in with our findings of section~3.1.

\subsubsection{The case $c=24$}

The case $c=24$ is actually the most interesting one from this point of view. At $c=24$
we are looking for modular forms of weight $12$. At genus one (degree one), the space
of modular forms is $2$-dimensional and we can take $\phi_4^3=\Theta_{e8}^3$
and $\Delta$ as generators. The condition $\Phi(W_1)=1$ implies then
\be
W_1=\phi_4^3+a\Delta\ ,
\ee
where $a$ is some constant (that will depend on the theory). The
corresponding partition function $\chi_1$ then is
\be
\chi_1=\frac{q}{\Delta}(\phi_4^3+a\Delta)
=q(j(\tau)+a)=1+(744+a)q+196884q^2+\ldots\ . \ee The coefficient of
$q$ in this expansion is the number $N$ of currents of the theory,
so that the genus $1$ partition function depends only on $N$
\be
W_1=W_1(N)=\phi_4^3+(N-744)\Delta\ ,\qquad \chi_1=q(j+N-744)\ . \ee
Let us consider the genus $2$ partition function. At grade $g=2$ the
space of modular forms of weight $12$ is $3$-dimensional, and it is
convenient to write $W_2$ as a linear combination of $\psi_4^3$,
$\chi_{12}$ and $\psi_{12}$. 
The condition
$\Phi(W_2)=W_1$ now implies that
\be
W_2=\psi_4^3+(N-744)\psi_{12}+b\chi_{12}\ , \ee for some constant
$b$. Next we impose the factorisation condition $W_2\to W_1\otimes
W_1$. Since
\be
W_1\otimes
W_1=\phi_4^3\otimes\phi_4^3+
(N-744)(\phi_4^3\otimes\Delta+\Delta\otimes\phi_4^3)+(N-744)^2\Delta\otimes\Delta
\ee
and since
\be
W_2\to \phi_4^3\otimes\phi_4^3+(N-744)(\phi_4^3\otimes\Delta+
\Delta\otimes\phi_4^3-1728\Delta\otimes\Delta)+b\Delta\otimes\Delta\
, \ee we obtain $b=(N-744)(N+984)$. Thus we find that
\be
W_2=W_2(N)=\psi_4^3+(N-744)\psi_{12}+(N-744)(N+984)\chi_{12} \ , \ee
and thus also the genus $2$ partition function is completely
determined by the number of currents. A similar result has also been
recently obtained in \cite{Tuite:1999id,Mason:2007ph}, using a
different approach.

The computation at genus $3$ is analogous. The modular form $W_3$ is
a linear combination of $\alpha_4^3$,  $\alpha_{12}$,
$\tilde\alpha_{12}$ and $\beta_{12}$, and the constraints are
$\Phi(W_3)=W_2$ and $W_3\to W_2\otimes W_1$. The first condition
gives
\be
W_3=\alpha_4^3+(N-744)\tilde\alpha_{12}+(N-744)(N+984)\alpha_{12}
+c\beta_{12}\ , \ee whereas the second one fixes
$c=(N+984)(N-744)^2$, so that
\begin{equation}\label{e:W3c24}
W_3=W_3(N)=\alpha_4^3+(N-744)\Bigl[\tilde\alpha_{12}+
(N+984)\bigl(\alpha_{12} +(N-744)\beta_{12}\bigr)\Bigr]\ .
\end{equation}
This proves our claim that the partition functions for $g\leq 3$ at $c=24$ are
uniquely determined in terms of the number of currents.

It is amusing to observe that the partition function $\chi_g$, for genus
$g=1,2,3$, has a polynomial dependence on the number of currents $N$, with the
degree of the polynomial being $g$. Following our general discussion, this therefore
implies that the expressions $\Tr_{\CFT_2}(C_2^l)$, for $l=0,1,2$, must have an analogous
polynomial dependence on $N$, with degree (at most) $l+1$. This holds trivially
for the case of $l=0$, since the dimension of $\CFT_2$ does not depend on $N$, as the
explicit expression for $\chi_1$ shows. For $l=1$ and $l=2$, however, this is a non-trivial
claim. By considering a few different theories, one can determine the coefficients of
the polynomials explicitly, and one finds
\begin{align}
\Tr_{\CFT_2}(C_2)&=-2 N^2+32808 N\ ,\\
\Tr_{\CFT_2}(C_2^2)&=-\frac{23 N^3}{36}+\frac{16421 N^2}{3}+40 N\ .
\end{align}
One can then check that these identities are in fact satisfied by {\em all}
meromorphic conformal field theories with central charge
$c=24$. This provides a highly non-trivial cross-check of the
correctness of the analysis in this section and of the results of
section \ref{s:results}.

\subsubsection{The case $c=32$}

For theories with central charge $c=32$, the space of modular forms of grade $1$
is still $2$-dimensional, and we may take the generators to be  $\phi_4^4$ and
$\Delta\phi_4$. It follows that, in this case,
\be
W_1=\phi_4(\phi_4^3+(N-992)\Delta)\ ,
\ee
and the genus $1$ partition function still depends only on $N$
\be
\chi_1=q\bigl(j(\tau)+(N-992)\bigr)\, \chi_1(E_8)= 1+N q+(248
N+139504) q^2+\ldots  \ . \ee At genus $2$, the space of modular
forms is generated by $\chi_{12}\psi_4$, $\chi_{10}\psi_6$,
$\psi_6^2\psi_4$, and $\psi_4^4$. Since $\chi_{10}$ is a cusp form
and vanishes when the period matrix is block diagonal, the
coefficient of $\chi_{10}\psi_6$ is not determined by factorisation
constraints. This implies that, in general, a pair of conformal
field theories of central charge $32$, with the same partition
function at $g=1$, may have a different partition function at genus
$2$. This is very nicely consistent with the explicit computations
of  section~3.

\subsection{Comments about genus $g\geq 4$}

The above analysis cannot be generalised to genus $g>3$ in a
straightforward manner. First of all, for $g\geq 4$, the closure
$\bar \J_g$ of the locus of Riemann period matrices does not
correspond to the whole Siegel upper half-space $\Sieg_g$ any
longer. This implies that $W_g$ does not necessarily extend to a
well-defined Siegel modular form on $\Sieg_g$. The second issue is
that the complete classification of Siegel modular forms of degree
$g>3$ is not known. For these reasons, a general treatment is not
possible for genera $g>3$. However, some results can be obtained for
the genus $g=4$ partition functions of lattice theories with central
charge $c\le 24$.

For $c=16$ and $g=4$, the theta series $\Theta_{e8}^2$ and
$\Theta_{d{16}}$ are distinct modular forms on $\Sieg_g$, but their
difference vanishes on $\bar \J_4$. Remarkably,
\be
J_8:=\Theta_{d{16}}-\Theta_{e8}^2=0\ , \ee is in fact the defining
equation for $\bar \J_4$ in $\Sieg_4$, thus providing the explicit
solution for the Schottky problem at $g=4$ \cite{Igusa:1979}. In
particular, any modular form vanishing on $\bar \J_4$ must be the
product of a modular form times some power of $J_8$.

For lattice theories at  $c=24$, $W_4$ must lie in the subspace of modular forms of
degree $4$ generated by theta series. Because of (\ref{Wconst}) the image of
$W_4$ under the Siegel operator must be given by $W_3(N)$ of
eq.~\eqref{e:W3c24}, where $N$ takes all the possible values in table
\ref{t:Niemeier}. It is easy to see from this expression that
the space generated by the different $W_3(N)$ (where $N$ attains all the different
allowed values) is actually $4$-dimensional. In particular, this shows that the whole
space of modular forms of degree $3$ and weight $12$ is generated by theta series.
This is true also for modular forms of degree $4$ and weight $12$
\cite{Bocherer:1983}. Furthermore, it is known that the space of cusp forms of
degree $4$ and weight $12$ is two dimensional \cite{Poor:1996}. One
such cusp form is $\Theta_{e8}J_8$, because
\be
\Phi(\Theta_{e8}J_8)=\Phi(\Theta^{(4)}_{e8\,
d{16}}-\Theta^{(4)}_{e8^3})= \Theta^{(3)}_{e8\,
d16}-\Theta^{(3)}_{e8^3}=0\ . \ee It then follows that the space of
modular forms of degree $4$ and weight $12$ is $6$-dimensional, and
we can choose a basis to consist of $\Theta_{E_8}J_8$, $K$, $\xi_4$,
 $\xi_{12}$, $\tilde \xi_{12}$ and $\rho_{12}$, where $K$ is a cusp form
and
\be
\Phi(\xi_4)=\alpha_4^3\ ,\quad \Phi(\xi_{12})=\alpha_{12}\ ,\quad
\Phi(\tilde\xi_{12})=\tilde\alpha_{12}\
,\quad\Phi(\rho_{12})=\beta_{12}\ ,\quad\ . \ee Then, the theta
series of degree $4$ can be written as
$$\Theta_{\Lambda}^{(4)}=c_4(N)\xi_4+c_{12}(N)\xi_{12}+
\tilde c_{12}(N)\tilde\xi_{12}+d_{12}(N)\rho_{12}+e\Theta_{E_8}J_8+f
K\ ,
$$
for some coefficients $c_4(N)$, $c_{12}(N)$, $\tilde c_{12}(N)$,
$d_{12}(N)$, $e$ and $f$, where $e$ and $f$ in principle depend on
$\Lambda$. In fact, the $c_4(N),c_{12}(N),\tilde
c_{12}(N),d_{12}(N)$ are uniquely fixed by the condition that
\be \Phi(\Theta_{\Lambda}^{(4)})=\Theta_{\Lambda}^{(3)}= W_3\ , \ee
{\it i.e.}\ they simply agree with the coefficients of $\alpha_4^3$,
 $\alpha_{12}$, $\tilde\alpha_{12}$ and $\beta_{12}$ in $W_3(N)$. Note that
all these coefficients are polynomials of degree at most $3$ in $N$.
In the limit $\Omega\to\Omega_{k,4-k}$, $k=1,2$, the theta series
satisfy the factorisation conditions
\begin{equation}\label{e:g4factor}\Theta_{\Lambda}^{(4)}\to
\Theta_{\Lambda}^{(k)}\otimes\Theta_{\Lambda}^{(4-k)}\ , \qquad k=1,2 \ .
\end{equation}
It is easy to see that, for both $k=1$ and $k=2$
\begin{equation}\label{e:factorJ}
\Theta_{E_8}J_8\equiv
\Theta^{(4)}_{e8\,d16}-\Theta^{(4)}_{(e8)^3}\to
\Theta^{(k)}_{e8\,d16}\otimes\Theta^{(g-k)}_{e8\,d16}-
\Theta^{(k)}_{(e8)^3}\otimes\Theta^{(g-k)}_{(e8)^3}=0\ .
\end{equation}
We now want to argue that the corresponding factorisation limit of
$K$ cannot be trivial. To see this we note that
$\Theta_{\Lambda}^{(k)}=W_k(N)$ for $k=1,2,3$, is a polynomial of
degree $k$ in $N$. Thus  $\Theta_{\Lambda}^{(k)}\otimes
\Theta_{\Lambda}^{(4-k)}$ is a polynomial of degree $4$. On the
other hand, the coefficients $c_4(N)$, $c_{12}(N)$, $\tilde
c_{12}(N)$, $d_{12}(N)$ are all polynomials of degree at most $3$.
If the factorisation limit of $K$ was trivial, the factorisation
constraint would lead to an identity between a polynomial of degree
at most $3$, and a polynomial of degree $4$. However, such an an
identity can at most be true for five different values of $N$. But
there are $19$ possible values for $N$ in table~\ref{t:Niemeier},
and it is thus impossible that the identity is true for all of them.
It therefore follows that the factorisation limit of $K$ is
non-trivial.

But if the factorisation limit of $K$ is non-trivial, then we can
determine the coefficient of $K$ via factorisation. By the same
argument as above, the coefficient of $K$ is then a polynomial in
$N$ of degree $4$. But since $\Theta_{e8}J_8$ vanishes on $\bar \J$,
this proves that the restrictions of the theta series of degree
$g=4$ to $\bar \J_4$ depends only on the number of currents, and
that the dependency is polynomial of degree $4$.

As in the lower genus case, such an analysis implies that the traces
$\Tr_{\CFT_2}(C_2^3)$ must be polynomial of degree $4$ in the number of
currents $N$. Again, we can fix the precise coefficients by comparison with
a few explicit examples, and we find that
\be
\Tr_{\CFT_2}(C_2^3)=-\frac{133 N^4}{864}+\frac{10969 N^3}{12}+2 N^2-272 N\ .
\ee
It is then again a non-trivial consistency check that the identity also holds
for the other Niemeier lattice theories at $c=24$. In fact, the identity actually holds
for all known $c=24$ theories; this suggests that the above results may be
more generally correct.
\smallskip

The same argument does not work at genus $g=5$, since at $g=5$
there exists a Siegel modular form $M$ of weight $12$ that does not vanish on the
moduli space of Riemann surfaces, but for which $\Phi(M) = \Theta_{e8} \, J_8$.
The coefficient of $M$ thus cannot be determined by factorisation arguments,
and will therefore depend on the actual structure of the theory. This is obviously
in perfect agreement with what we saw explicitly in our analysis of section~3.

\section{A general approach}\label{s:geninvariants}

The analysis of the previous section suggests that one should be
able to identify the Lie symmetry of a given conformal field theory from
its genus $g$ vacuum amplitudes. We now want to show that this is indeed so.
A convenient method to approach this problem is to
consider more general degeneration limits
of genus $g$ surfaces.

\subsection{Invariants from partition functions}\label{s:invconst}

\begin{figure}[htb]\begin{center}
\resizebox{.8\textwidth}{!}{\input{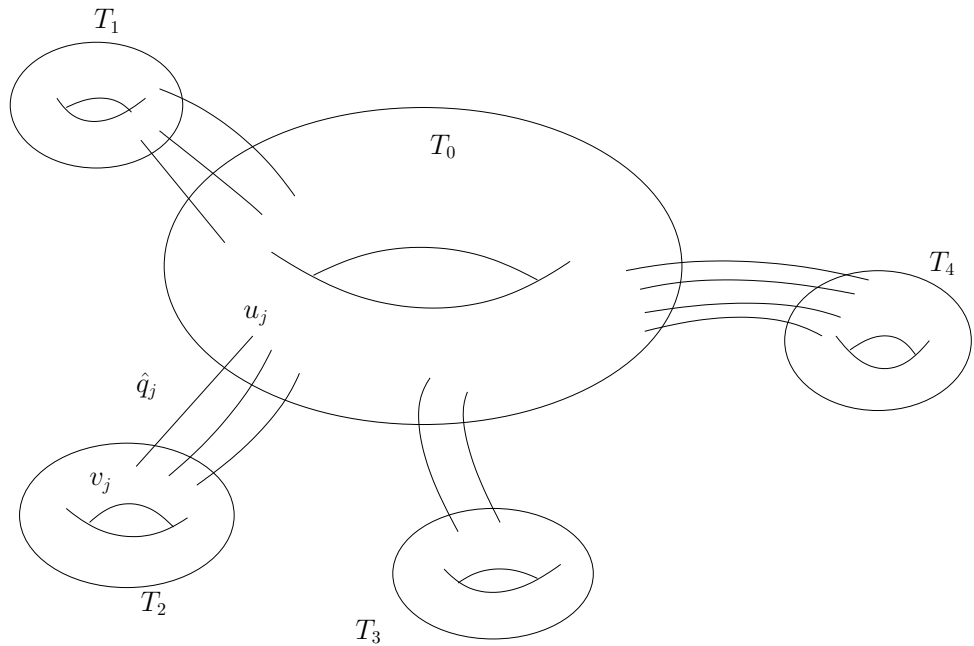}}\end{center}
\caption{\small A singular Riemann surface of genus $12$,
corresponding to $r=4$, $l_1=3$, $l_2=3$, $l_3=2$, $l_4=4$. Each line
represents a node connecting a torus $T_i$ to the torus
$T_0$.}\label{f:lotstori}\end{figure}

Given a genus $g$ Riemann surface we want to consider the degeneration limit
that is sketched in figure~\ref{f:lotstori}.
Its connected components, once the nodes are removed, are $r+1$ tori
$T_0,T_1,\ldots,T_r$, with modular parameters $q_0,q_1,\ldots,q_r$.
The torus $T_j$ is connected by $l_j$ nodes to the torus $T_0$, but there
are no nodes connecting two tori $T_j$ and $T_k$ with $k,j>0$, and no nodes
identifying distinct points on the same torus. Thus, the total number of nodes is
$n=\sum_j^r l_j$, and each of them is associated with a degeneration
parameter $\hat{q}_j$ and two points $u_j,v_j$, $j=1,\ldots,n$, with
$u_j\in T_0$ and $v_j$ on a torus $T_i$ for some $i>0$. The
genus $g$ of such a singular Riemann surface can be read off
directly from the geometrical sketch: each torus $T_j$ with the $l_j$ connecting
nodes adds $l_j$ handles; together with the torus $T_0$ in the middle, the total genus is
therefore $g=n+1$. This also ties in with the counting of the moduli:
there are $3$ parameters $\hat{q}_j$, $u_j$, $v_j$ associated to each node, and
each torus has one modular parameter and one symmetry, so that the total number of
independent parameters is $3n$. A surface of genus $g>1$ has $3g-3$ moduli, so
that this also gives $g=n+1$.

Let us consider the expansion of the genus $g=n+1$ character of a
meromorphic CFT in the limit $\hat{q}_1,\ldots,\hat{q}_n\to 0$. The coefficient
of the term  $\prod_j \hat{q}_j$ is given by a product of $r$
correlation functions of currents, one per torus,
\begin{equation}
\left. \chi_g \right|_{\prod_{j=1}^n \hat{q}_j} =
\Tr_\CFT\Bigl( q_0^{L_0}\prod_{i=1}^n\hat{J}^{a_i}(u_i)\Bigr)\,
\Tr_\CFT\Bigl( q_1^{L_0}\prod_{i=1}^{l_1}\hat{J}^{a_i}(v_i)\Bigr)
\cdots
\Tr_\CFT\Bigl( q_r^{L_0}\prod_{i}\hat J^{a_{i}}(v_i)\Bigr)\ .
\end{equation}
The indices of the $l_j$ currents appearing in the correlator on the
torus $T_j$, $j>0$, are contracted with a set of $l_j$ currents in
the correlator on $T_0$. By integrating all the points $u_j,v_j$
around the $\alpha$-cycles of the respective tori, we pick up the zero
modes of the currents and obtain a product of traces
\begin{equation}\label{traceprod}
\Tr_{\CFT}\bigl(q_0^{L_0}\hat J^{a_1}_0\cdots \hat J^{a_n}_0\bigr)\,
\Tr_{\CFT}\bigl(q_1^{L_0}\hat J^{a_1}_0\cdots \hat J^{a_{l_1}}_0\bigr) \,
\cdots  \,
\Tr_{\CFT}\bigl(q_r^{L_0}\hat J^{a_{n-l_r+1}}_0\cdots \hat J^{a_{n}}_0\bigr)\ .
\end{equation}
Here we have picked some particular order for the integration paths
of the points; this is not the most general case (and indeed we could
consider more complicated degenerations, for example when there are also
nodes between $T_i$ and $T_j$ with $i,j>0$), but  for our present purposes,
this will suffice.

The product of traces in (\ref{traceprod}) can be expanded in powers of
$q_0,\ldots,q_r$, and the coefficient of the term
$q_0^{h}q_1^{h_1}\cdots q_r^{h_r}$ is
\be\label{e:trHH}
\Tr_{\CFT_{h}}\bigl(\hat t^{a_1}\cdots \hat t^{a_n}\bigr)\,
\Tr_{\CFT_{h_1}}\bigl(\hat t^{a_1}\cdots \hat t^{a_{l_1}}\bigr)\cdots
\Tr_{\CFT_{h_r}}\bigl(\hat t^{a_{n-l_r+1}}\cdots \hat t^{a_{n}}\bigr)\ , \ee
where we denote by $\hat t^a$ the rescaled Lie algebra generators
(compare (\ref{Jres}))
\begin{equation}
\hat t^a= k^{-\frac{1}{2}}\, t^a \qquad \hbox{with} \quad
t^a \equiv J^a_0 \ ,
\end{equation}
and $k$ is the level of the corresponding Lie algebra. For the following
it is convenient to define the Casimir operators of degree $l$ (see
for example \cite{Knapp})
\be\label{e:AdCas}
C_l^{(\g)}:=\Tr_{ad}(t^{a_1}\cdots t^{a_l})\,
t^{a_1}\cdots t^{a_l}\ , \qquad l=2,3,\ldots \ ,
\ee
where we sum over an orthonormal basis with respect to the Killing form (see
(\ref{Lienorm})), and $ad$ denotes the adjoint representation of $\g$. For example,
for $l=2$, this is just the rescaled quadratic Casimir operator
\be
C_2^{(\g)} = 2 \, h^\vee \, C_2 \ , \qquad \hbox{since} \qquad
\Tr_{ad} (t^a t^b) = 2 \, h^\vee \, \delta^{ab} \ .
\ee
In terms of these Casimir operators we can then express \eqref{e:trHH}
for $h_1=\cdots = h_r =1$, with $h$ being arbitrary, as
\be\label{e:trAdH}
k^{-n}\Tr_{\CFT_{h}} \bigl(C_{l_1}^{(\g)}\, C_{l_2}^{(\g)}\cdots C_{l_r}^{(\g)}\bigr) \ .
\ee
Here we have assumed that the Lie algebra $\g$ is simple, so that there is only
one level $k$; in general, if $\g = \oplus\ \g_i$, where $\g_i$ has level $k_i$, we
get instead of (\ref{e:trAdH})
\be\label{e:trAdHg}
\Tr_{\CFT_h}
\left[ \, \prod_{j=1}^r
\left(\sum_i \, k_i^{-l_j} C_{l_j}^{(\g_i)} \right) \right] \ .
\ee
Note that the trace $\Tr_{\CFT_1}(t^{a_p} \cdots t^{a_{p+l_j}})$
is only non-zero if all generators $t^{a_s}$, $s=p,\ldots,p+l_j$
lie in the same simple Lie algebra $\g_i$.

\subsection{Identifying the Lie algebra}
\label{s:5.2}

In the following we want to show that one can determine the affine
Lie algebra from suitable degeneration limits of higher genus partition
functions. The Lie algebra generators appear at $h=1$, and thus we should
consider (\ref{e:trHH}) (or (\ref{e:trAdH}) and (\ref{e:trAdHg})) for $h=1$.
Let us denote the rescaled value of the Casimir operator $C_l^{(\g)}$
in the adjoint representation $ad(\g)$ by
\be
\xi_l(\g,k) = \frac{C_l^{(\g)}(ad(\g))}{k^l} \ .
\ee
If the affine algebra is a direct sum of simple affine Lie algebras (and $\hat{u}(1)$ factors),
$\hat\g = \oplus\, n_i\, \hat\g_i$, where $\hat\g_i$ has
level $k_i$ and the $n_i$ are multiplicities, then (\ref{e:trAdHg}) becomes simply
\be
\sum_i \, k_i^{-n} \Tr_{\CFT_1}
\bigl( C_{l_1}^{(\g_i)}\, C_{l_2}^{(\g_i)}\cdots C_{l_r}^{(\g_i)}\bigr)  =
\sum_i n_i\, \dim(\g_i) \, \prod_{j=1}^{r} \xi_{l_j}(\g_i,k_i) \ .
\ee
By taking linear combinations of such invariants we can obtain
any polynomial of the $\xi_l(\g_i,k_i)$, {\it i.e.}\ we can get expressions for
\be\label{poly}
\sum_i n_i \, \dim(\g_i) \, P\bigl(\xi_{2}(\g_i,k_i), \xi_{3}(\g_i,k_i),\ldots\bigr) \ ,
\ee
where $P$ is an arbitrary polynomial. In fact, the vacuum amplitudes up to genus
$g$ gives us access to all polynomials whose total degree is $g-1$
(where we regard $\xi_l(\g_i,k_i)$ as having degree $l$).
\smallskip

The main strategy for our argument is now as follows. Since the
dimension of $\CFT_1$ is finite, it is clear that only finitely many possible
$\g_i$ may appear in $\g$. We can also show (see section~\ref{s:kbound} below
for the detailed argument) that only finitely many values of $k_i$ are possible. Thus
there are only finitely many possibilities for $\hat\g_i$ we have to distinguish.

The second ingredient is that any simple affine algebra
$\hat\g$ at level $k$ is uniquely identified by its values for $\xi_l$. More specifically,
as shown in more detail below in section~\ref{s:Qtrick}, for any pair of
simple affine Lie algebras $\g_i$ at level $k_i$ and $\g_{j}$  at level $k_j$,
for which either $\g_i\neq \g_j$ or $k_i\neq k_j$, there exists an $2\leq l_{ij} < \infty$
such that
\be\label{ineq}
\xi_{l_{ij}}(\g_i, k_i)\neq \xi_{l_{ij}}(\g_j, k_j) \ .
\ee
Then we can consider the polynomial
\be
P_i(x_2,x_3,\ldots) = \prod_{j\neq i}
\frac{x_{l_{ij}} - \xi_{l_{ij}}(\g_j,k_j)}{\xi_{l_{ij}}(\g_i, k_i) - \xi_{l_{ij}}(\g_j, k_j)}  \ ,
\ee
where $j$ runs over all the finitely many possibilities for $\hat\g_j$. If
we apply (\ref{poly}) with $P=P_i$, then we simply obtain $n_i \dim(\g_i)$.
This allows us to read off the multiplicity with which $\hat\g_i$ appears
in $\hat\g$.

Since (\ref{poly}) with $P=P_i$ can be obtained from a suitable
degeneration limit of the vacuum genus $g$ amplitudes (where $g$ is sufficiently
large such that the degree of all $P_i$ is less than $g-1$),
this argument allows us to identify $\hat\g$ uniquely. Put differently, if
two meromorphic conformal field theories contain different affine algebras,
then their vaccum amplitudes cannot agree for all genera.
\smallskip

As an example, let us consider the $E_8\times E_8$ and
$Spin(32)/\ZZ_2$ theories. The dual Coxeter numbers are the same, so
that $C_2^{(\g)}(ad(\g))$ is the same for both theories. However, the two
Lie algebras have also a fourth order Casimir $C_4^{(\g)}$, which
can be obtained from a genus $5$ partition function. In the adjoint
representations it equals (the details of this computations are explained in
section~\ref{s:Qtrick})
\begin{equation}\label{e:quartic}
\begin{array} {rl}
E_8\times E_8:
&\Tr_{\CFT_1}(t^{a_1}\cdots t^{a_4})\Tr_{\CFT_1}(t^{a_1}\cdots
t^{a_4})=2\dim(e8)\, C_4^{(\g)}(ad(\g))=589248000\\
Spin(32)/\ZZ_2:
&\Tr_{\CFT_1}(t^{a_1}\cdots t^{a_4})
\Tr_{\CFT_1}(t^{a_1}\cdots t^{a_4}) =\dim(d16)\, C_4^{(\g)}(ad(\g))=749237760,
\end{array}
\end{equation}
and hence allows one to distinguish the two theories at genus $g=5$,
in agreement with the earlier analysis.
\smallskip

In order to complete our argument it remains to explain the two
remaining issues, namely (i) that there are only finitely many possible
affine algebras that may appear; and (ii) that (\ref{ineq}) holds. We shall
first deal with (i).

\subsubsection{The bound on the level}\label{s:kbound}

Since
\be
\dim(\CFT_1) = \sum_i n_i\, \dim(\g_i)
\ee
it is clear that only those Lie algebras $\g_i$ may appear in $\g$ that
satisfy $\dim(\g_i)\le\dim(\CFT_1)$. Given $\dim(\CFT_1)$, there are
therefore only finitely many possibilities for $\g_i$. However, this dimensional
reasoning does not give a constraint on the possible levels $k_i$. In
this section, we will show that the levels are also bounded.

\noindent The starting point of our analysis is the quantity
\be\label{e:a12}
A:=\Tr_{\CFT_1}(\hat t^a\hat t^b)\Tr_{\CFT_1}(\hat t^a\hat t^b)
=\sum_i \frac{\dim(\g_i)\, C^{(\g_i)}_2(ad(\g_i))}{k_i^2} \ ,
\ee
that may be obtained from the degeneration of the genus $g=2$
vacuum amplitude. By virtue of (\ref{e:a12}) $A$ is a rational number.
We can thus find a positive integer $M$ such that $AM\in\NN$, as well as
\be
x_i:=M\dim(\g_i)\, C^{(\g_i)}_2(ad(\g_i))\in\NN
\ee
for all $i$ with $\dim(\g_i)\leq \dim(\CFT_1)$.
By multiplying both sides of \eqref{e:a12} by $M$ we then obtain
\be\label{e:sumxk}
\sum_i \frac{x_i}{k_i^2}=MA\in\NN\ .
\ee
Note that the numerators $x_i$ are uniformly bounded
\be\label{e:boux}
x_i\le X\ ,
\ee
for some $X$, because each $x_i$ only depends on the Lie algebra $\g_i$
as well as the choice of $M$.

\noindent Let $k_1$ be the smallest level that appears in $\hat\g = \oplus n_i\, \hat\g_i$.
The right hand side of \eqref{e:sumxk} is a positive integer, and hence must at least be
equal to $1$. On the other hand, the left hand side is a sum over at most
$N=\dim(\CFT_1)$ positive terms, each of which is bounded by
\be
\frac{x_i}{k_i^2} \leq \frac{X}{k_1^2} \ .
\ee
It therefore follows that
\be
N\frac{X}{ k_1^2}\ge 1\ ,
\ee
and hence $k_1$ is bounded by
\be\label{e:boundk1}
k_1^2\le XN\ .
\ee
If $k_1$ is the only level appearing in the decomposition of $\hat\g$, we are done.
Otherwise let us multiply both sides of eq.~\eqref{e:sumxk} by $k_1^2$ to obtain
\be
\sum_{i\ge 2}
\frac{k_1^2\, x_i}{k_i^2}= k_1^2\, MA -x_1\in \NN \ .
\ee
We choose our numbering such that $k_2$ is the second smallest level. Then we
repeat the argument where now the numerators $k_1^2x_i$ are uniformly bounded
by  $X^2N$. Since the right hand side is still positive, we thus obtain the inequality
\be
(N-1)\frac{X^2N}{k_2^2}\ge 1\qquad \Rightarrow \qquad
k_2^2\le X^2N(N-1)\ .
\ee
Repeating this procedure (at most $N$ times) we obtain an upper bound
for all possible levels $k_i$ appearing in the decomposition of $\hat\g$.

\subsubsection{Higher degree Casimir invariants in the adjoint
representation}\label{s:Qtrick}

Thus it only remains to prove \eqref{ineq} for any pair of
affine Lie algebras $\g$ at level $k$ and $\g'$ at level $k'$ for which
either $\g \neq \g'$ or $k\neq k'$. Given a simple Lie algebra $\g$, consider
the linear operator  $Q$ acting on the tensor product representation
$ad\otimes ad$ \cite{Okubo:1977mx}
\be
Q=\sum_a t^a \otimes t^a\ ,
\ee
where $t^a$ acts in the standard way on $ad$. The trace
of its $l$'th power equals
\be\label{e:qltrace}
\Tr_{ad\otimes ad}(Q^l)=\sum_{a_1,\ldots,a_l}\Tr_{ad}(t^{a_1}\ldots
t^{a_l})\Tr_{ad}(t^{a_1}\ldots t^{a_l})
=\dim(\g)\, C^{(\g)}_l(ad(\g))\ .
\ee
The Lie algebra generators $t^a$ act on the tensor product $ad\otimes ad$ as
\be
y^a=t^a \otimes {\bf 1}+{\bf 1}\otimes t^a\ .
\ee
In terms of these generators we can write the operator $Q$ as
\be
Q=\frac{1}{2}\sum_a\Bigl(y^ay^a-(t^at^a \otimes{\bf 1})
- ({\bf 1}\otimes t^at^a)\Bigr)\ .
\ee
Let $ad\otimes ad=\oplus_iR_i$ be the decomposition of the tensor
product $ad\otimes ad$ into irreducible representations, and let $P_i$ be the
projector onto $R_i$. Then we have
\be
\sum_ay^ay^a=\sum_iC_2(R_i)P_i\ ,\qquad
\sum_a(t^at^a \otimes {\bf 1})=\sum_a ({\bf 1}\otimes t^at^a)=C_2(ad) ({\bf 1}\otimes {\bf 1})
\ee
and hence $Q=\sum_i \lambda_i\, P_i$, where
\be\label{lambdai}
\lambda_i=\frac{C_2(R_i)}{2}-C_2(ad)
\ee
are the eigenvalues of $Q$, so that
\be\label{e:Qcas}
C^{(\g)}_l(ad(\g))=\frac{\Tr_{ad\otimes ad}(Q^l)}{\dim(\g)}
=\sum_i\frac{\dim(R_i)}{\dim(\g)}\, \lambda_i^l\ .
\ee The eigenvalues of $Q$ for all the simple Lie algebras are
listed in table \ref{t:algeigen}.

\begin{table}[htb]\begin{center}
\begin{tabular}{|c|rrrrrr|}\hline Algebra & \multicolumn{6}{|l|}{Eigenvalues}\\
\hline $u(1)$ & 0 & & & & &  \\
$a1$ & $-4$ & $-2$ & $2$ &&&\\
$a2$ & $-6$ & $-3$ & $0$ & $2$ &&\\
$a(r)$, $r>2$ & $-2(r+1)$ & $-(r+1)$ & $-2$ & $0$ & $2$ &  \\
$b3$ & $-10$ & $-5$ & $-4$ & $-3$ & $0$ & $2$\\
$b(r)$, $r> 3$ & $-2(2r-1)$ & $-2r+1$ & $-2r+3$ & $-4$ & $0$ & $2$\\
$c(r)$, $r\ge 2$ & $-2(r+1)$ & $-(r+2)$ & $-(r+1)$ & $-1$ & $0$ & $2$\\
$d4$& $-12$ & $-6$ & $-4$ & $0$ & $2$&\\
$d(r)$, $r> 4$ & $-4(r-1)$ & $-2r+2$ & $-2r+4$ & $-4$ & $0$ & $2$\\
$e6$ & -24 & -12 & -6 & 0 & 2 & \\
$e7$ & -36& -18  & -8 & 0 & 2& \\
$e8$ & -60& -30  & -12& 0 & 2 & \\
$f4$ & -18 & -9 & -5 & 0 & 2& \\
$g2$ & -8 & -4 & $-\frac{10}{3}$ & 0& 2&\\
\hline\end{tabular}\caption{\small The different eigenvalues
$\lambda_i$ (\ref{lambdai}) of $Q$ for all the simple Lie algebras
and for $u(1)$. In each row the eigenvalues are given in increasing
order.}\label{t:algeigen}
\end{center}\end{table}

\noindent For example for the $e8$ and $d16$ algebras, the operator $Q$ equals
\begin{align*}&e8:& Q&=-60\cdot P_{\bf 1}-30\cdot P_{\bf 248} -12\cdot P_{\bf 3875}
+0\cdot P_{\bf 30380} +2\cdot P_{\bf 27000}\ ,\\
&d16: & Q&=-60\cdot P_{\bf 1}-30\cdot P_{\bf 496}
-28\cdot P_{\bf 527}-4\cdot P_{\bf 35960}+0\cdot P_{\bf 122264}
+2\cdot P_{\bf 86768}\ ,
\end{align*}
where we have labelled the different projectors $P_i$ by the dimension of $R_i$.
Together with the equations \eqref{e:qltrace} and \eqref{e:Qcas}  this
then leads to \eqref{e:quartic}.
\smallskip

Now we can prove our claim (\ref{ineq}). If $\g=u(1)$, then all $\xi_l(\g,k)=0$,
and thus also $\g'=u(1)$. Otherwise, if $\xi_l(\g,k)=\xi_l(\g',k')$ for
all $l$, then this implies, because of (\ref{e:Qcas}), that all
the eigenvalues of $Q/k$ and $Q'/k'$ must agree --- the factors $\dim(\g)$ and
$\dim(\g')$ only affect the multiplicities of such eigenvalues. But this then implies
that all the eigenvalues $\lambda_i$ of $Q$ and $\lambda'_i$ of $Q'$ must be
related as
\be\label{e:eigenv}
\lambda_i=\frac{k}{k'}\lambda'_i\ .
\ee
Each simple Lie algebra has a unique positive eigenvalue equal to $2$, and thus
(\ref{e:eigenv}) can only be satisfied if $k=k'$. But then (\ref{e:eigenv}) requires
that the eigenvalues of $Q$ and $Q'$ are the same, but it is immediate from
table~\ref{t:algeigen} that this is only possible if $\g=\g'$. Thus $\xi_l(\g,k)=\xi_l(\g',k')$
for all $l$ implies that $\g=\g'$ and $k=k'$. This completes our proof.

\subsection{Identifying representations}

In the previous section we have seen that we can determine the affine
algebra symmetry of a meromorphic conformal field theory from its vacuum
amplitudes. An obvious refinement of this question is whether we can similarly
determine the representation content of the theory.

To answer this question we proceed in the same manner as before, except that we
now take $h$ in (\ref{e:trAdHg}) to assume any value, not just $h=1$. Since $\dim(\CFT_h)$
is finite, only a finite set of irreducible representations of the Lie algebra $\g$
can appear in the decomposition $\CFT_h=\oplus_iR_i$. Furthermore, by
the same arguments as in section~\ref{s:5.2}, the vacuum amplitudes
determine the trace over $\CFT_h$ of any polynomial in the Casimir operators
$C_l^{(\g)}$. Using the same techniques as above, the question of whether we
can determine the representation content uniquely then boils down to the question
of whether we can distinguish all representations $R_i$ by their eigenvalues
with respect to the Casimir operators $C_l^{(\g)}$. In the following we shall assume
that $\g$ is simple; we shall come back to question of how to deal with
the semi-simple case in section~\ref{5.4}.

It is well known that we can distinguish the representations of any simple
Lie algebra $\g$ by the eigenvalues of {\em all invariants}. The algebra of invariants
of a simple Lie algebra $\g$ is generated by a set of  $\rank(\g)$ Casimir operators
\be
C^\perp_l:=c^{a_1\ldots a_l}t^{a_1}\ldots t^{a_l}\ ,
\ee
where $l$ takes values in a finite set of degrees that depends on the Lie algebra
$\g$ in question, and we are using again the orthonormal basis with respect to the
Killing form --- see (\ref{Lienorm}). The tensors $c^{a_1\ldots a_l}$ can be taken to be
totally symmetric in the indices, and to satisfy an orthonormality
condition
\be\label{ortho}
c^{a_1\ldots a_l}\, c^{a_1\ldots a_l}=1\ ,\qquad
c^{a_1\ldots a_l}\, c^{a_1\ldots a_{l'}}=0\ , \quad \hbox{if $l'\neq l$.}
\ee
The Casimir operators $C_l^{(\g)}$  we have used above (see (\ref{e:AdCas}))
can obviously be expressed in terms of these generators as
\be\label{Crel}
C_l^{(\g)}=I_l(\g)C_l^\perp+\text{polynomial in Casimirs of lower degree}\ .
\ee
Using the orthonormality condition (\ref{ortho}), the index $I_l(\g)$ turns out to be
\be
I_l(\g)=\Tr_{ad}(t^{a_1}\cdots t^{a_l})\, c^{a_1\ldots
a_l}=\dim(\g)\, C_l^\perp(ad(\g))\ . \ee This allows us to determine
the subalgebra generated by the $C_l^{(\g)}$ in principle.

It is not difficult to see that the Casimirs $C_l^{(\g)}$ agree on two
representations that are related to one another by an (outer)
automorphism of the Lie algebra. Thus it is clear that we cannot
distinguish between two representations that are related to one another
in this way. However, it is natural to conjecture (and we have circumstancial
evidence for it --- see appendix~C), that this is the only ambiguity:
\smallskip

\noindent {\bf Conjecture:} If  $R_1$ and $R_2$ are two irreducible representations
of a simple Lie algebra $\g$ such that the eigenvalues of $C_l^{(\g)}$ on $R_i$
are equal,
\be\label{condl}
C_l^{(g)} (R_1) =  C_l^{(g)} (R_2) \qquad
\hbox{for all $l$}
\ee
then either $R_1\cong R_2$ or $R_1\cong \pi(R_2)$, where $\pi$ is a
non-trivial (outer)  automorphism of $\g$.
\smallskip

For the simple Lie algebras (that we are currently considering) the only
non-trivial outer automorphisms are charge conjugation for $a(r)$, $e6$
and $d(r)$ with $r$ odd. For $d(r)$ with $r$ even, the outer automorphism
changes the chirality (spin flip) but does not map a representation to its conjugate.
Finally, there is the special case of $d4 = so(8)$, for which there is `triality'.

If the conjecture is true, then our analysis allows us (for $\g$ simple) to identify
the representation content at each conformal weight up these automorphisms.
Since the actual spectrum has to be real, we know on the other hand, that
all representations must appear in complex conjugate pairs. Thus the ambiguity
related to charge conjugation is irrelevant. The only genuine ambiguity then
occurs for the case of $d(n)$ with $n$ even, where our analysis does not let us
distinguish between representations of the opposite chirality; for $d(4)$ there is
in addition triality.

Obviously an {\em overall spin-flip} relates isomorphic conformal field theories to
one another, and we therefore should not be able to distinguish such theories.
However, on the basis of our present analysis we have not yet shown that the ambiguity
is just an overall spin-flip. In particular, we cannot yet distinguish between two
conformal field theories for which, say $\CFT^{(1)}_h = S_+ \oplus S_-$
and $\CFT^{(2)}_h = S_+ \oplus S_+$ for some $h$, where $S_\pm$ describe spinor
representations of opposite chirality. We shall come back to this point in
section~5.4.2.

\subsection{Other degeneration limits}\label{5.4}

\begin{figure}[htb]\begin{center}
\resizebox{\textwidth}{!}{\input{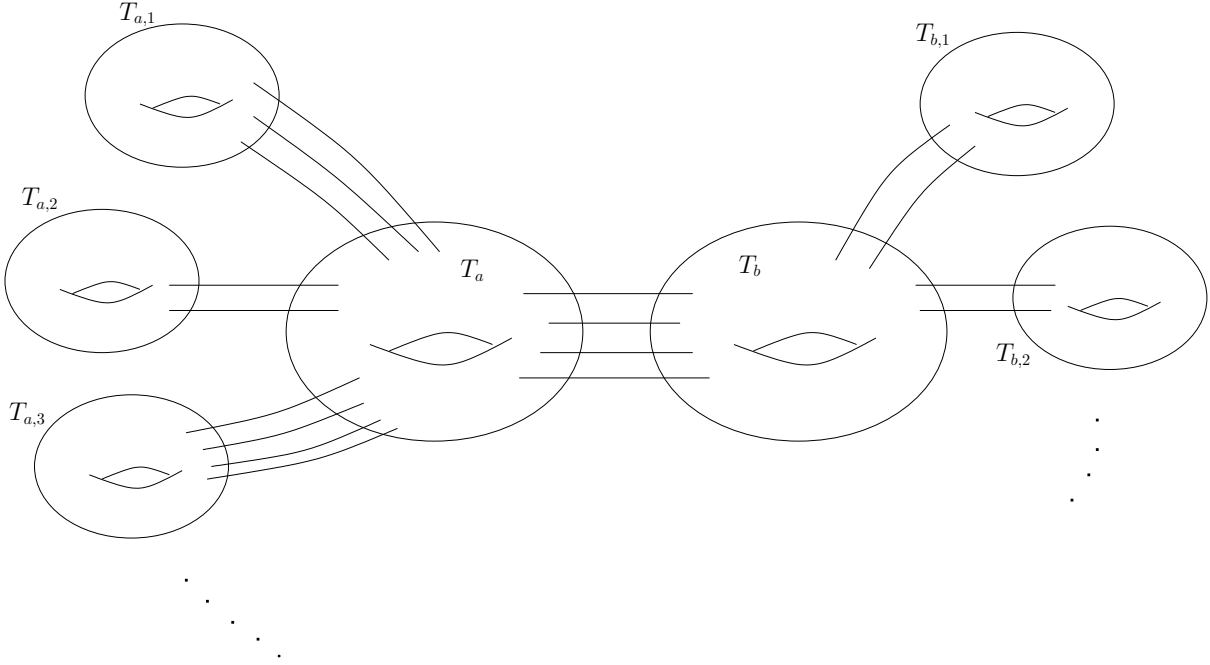}}\end{center}
\caption{\small A more general degeneration limit. We are interested
in the expansion where the modular parameters of the tori
$T_{a,1},T_{a,2},\ldots, T_{b,1},\ldots$ are taken to linear order,
while we consider the power $q_a^{h_a}\, q_b^{h_b}$ for the modular
parameters of the two tori $T_a$ and
$T_b$.}\label{f:toomanytori}\end{figure}

There are two issues that remain to be discussed: first the question of how
to deal with semi-simple Lie algebras (see the discussion at the beginning of section~5.3);
and secondly the question of how to show that the spin flip ambiguity is only
an overall ambiguity (see the end of previous secton). Both of these questions can
be addressed by considering more general degeneration limits of the type
depicted in figure~\ref{f:toomanytori}. We shall not attempt to develop the
general theory, but our arguments below will suggest how both problems
can be solved using such techniques.

\subsubsection{Direct sums of algebras}

Up to now we have implicitly discussed the case where $\g$ is a simple
affine algebra. The situation where $\hat\g = \oplus_i \, n_i \, \hat\g_i$ can
be dealt with similarly. Recall from section~\ref{s:geninvariants}
that we can define polynomial Lie algebra invariants
$P_{i}$ that act in $\CFT_1$ as a projector onto the subalgebra $\g_i$.
By taking $h_a=1$ with $h_b$ arbitrary, as well as the modular parameters of
the nodes between the tori $T_a$ and $T_b$ to be at linear order,
we can obtain from the above degeneration limit (see figure~\ref{f:toomanytori})
the invariant
\be\label{e:lotalgebras}
\Tr_{\CFT_1}(P_{i}\hat t^{a_1}\cdots \hat t^{a_l})\,
\Tr_{\CFT_h}(\hat t^{a_1}\cdots \hat t^{a_l})\ .
\ee
The first trace is only non-zero, if all $t^{a_j}$ lie in $\g_i$, and thus we can
identify the representation content with respect to this Lie algebra separately
from the rest. Using the techniques from the previous section, this allows us to
deal with the case where all $n_i=1$.

If some affine Lie algebra appears with higher multiplicity, the situation is
more complicated. However, this has to be so since theories with a non-trivial
multiplicity also have a bigger outer automorphism symmetry, namely
the permutation symmetry that exchanges the different copies of $\hat\g_i$.

\subsubsection{Spin flipped representations for $d(r)$ with $r$ even}\label{s:chiral}

As we explained
above, so far we cannot distinguish between theories
$\CFT_h = m_+\, S_+ \oplus m_- S_-$ with different values for $(m_+,m_-)$.  In fact,
the techniques of the previous section only allow us to determine
$m_+ + m_-$. We now want to show how we can also determine
$(m_+ - m_-)^2$. (We should not be able to determine directly $(m_+ - m_-)$
since the overall spin-flip exchanges $m_+$ and $m_-$ and hence
changes the sign of $(m_+ - m_-)$.)

To this end we now consider the degeneration limit of
figure~\ref{f:toomanytori} with $h_a=h_b=h$. Furthermore we consider
combinations of such configurations for which the external tori
($T_{a_1},T_{a_2},\ldots, T_{b_1},\ldots$) generate projectors $P_S$
onto $m_+\, S_+ \oplus m_- S_-$ --- this is possible since the Casimirs
$C_l^{(\g)}$ allow us to define such projectors. Thus we can obtain
the invariant
\be\label{e:spin1}
\Tr_{\CFT_h}(P_S\, t^{a_1}\cdots t^{a_r})\,
\Tr_{\CFT_h}(P_S\, t^{a_1}\cdots t^{a_r})\ .
\ee
This product of traces can be decomposed as
\be
\Tr_{\CFT_h}(P_S\, t^{a_1}\cdots t^{a_r})\,
\Tr_{\CFT_h}(P_S\, t^{a_1}\cdots t^{a_r})
=a\Tr_{\CFT_h}(P_S\, \tilde C_r^\perp)\,
\Tr_{\CFT_h}(P_S\, \tilde C_r^\perp)+\ldots\ ,
\ee
where $a$ is a non-zero coefficient which can be explicitly computed
and the ellipses denote the terms corresponding to polynomials of degree
$r$ in $C_l^{\perp}$ with $l<r$. These terms can be computed explicitly and
depend on the eigenvalues $C_l^{\g}(S^{\pm})$ and on the sum
$m_++m_-$ of the multiplicities. Thus we obtain
\be
\Tr_{\CFT_h}(P_S\, t^{a_1}\cdots t^{a_r})\,
\Tr_{\CFT_h}(P_S\, t^{a_1}\cdots t^{a_r})
=a\dim(S^{\pm})^2
\bigl|\tilde C_r^\perp(S^{\pm})\bigr|^2\, (m_+-m_-)^2
+\ldots\ .
\ee
Thus we can indeed determine $(m_+-m_-)^2$.

It should similarly be possible to determine the relative
chiralities at different conformal weights, simply by repeating the
argument for $h_a\neq h_b$. In this way one should be able to show
that the vacuum amplitudes allow one to identify these theories up
to an overall spin flip.

\section{Conclusions}
\setcounter{equation}{0}

In this paper we have studied the question of whether  a conformal
field theory is uniquely characterised by its higher genus vacuum amplitudes.
For the case of a meromorphic (chiral) conformal field theory we have shown that
the affine Lie algebra symmetry (that is generated by the currents at $h=1$) can be
determined uniquely from the higher genus vacuum amplitudes. We have
also given strong arguments that suggest that the vacuum amplitudes specify the
representation content of the theory (with respect to this affine algebra),
up to an overall automorphism of the finite Lie algebra.

We have applied our general arguments to some simple interesting examples,
in particular the self-dual theories at $c=16$ and $c=24$. Among other things
this has allowed us to give an elementary proof that the
$E_8\times E_8$ and the $Spin(32)/\mathbb{Z}_2$ theories at $c=16$ have different
genus $g=5$ vacuum amplitudes. The fact that the discrepancy only occurs at
a rather high genus is a consequence of the modular properties of higher genus
amplitudes at small values of the central charge. In particular, at $c\leq 24$ the
genus one amplitude already determines the amplitudes for genus $g\leq 4$
uniquely. On the other hand, at $c=32$, the different theories have typically already
different genus $g=2$ amplitudes.

For ease of notation we considered only meromorphic (chiral) theories in this paper.
It should be fairly obvious how to reformulate our arguments in the general case.
In particular, the analogue of (\ref{e:expand}) will in general be a power series in
$q_j^{h_j} \bar{q}_j^{\bar{h}_j}$, and we can thus pick out the contribution from the states
with arbitrary left- and right-moving conformal weights $(h_j,\bar{h}_j)$.
For example, in order to determine the left-moving affine symmetry, we can consider the terms
that go as $q_j^1 \bar{q}_j^0$, {\it etc.}, and the analysis is then essentially the
same as in the meromorphic context. Similarly, the representation content can be
determined with respect to both left- and right-moving affine algebras, up to
separate automorphisms of the left- and right-moving Lie algebra.
\smallskip

Our arguments thus go a certain way towards showing that a conformal field theory
is uniquely determined by its vacuum amplitudes. However, it should be clear
that they do not settle the question completely. In particular, we cannot say
much about theories without any current symmetries, such as for example the
Monster theory, although similar techniques will clearly also constrain
these theories. It would be interesting to gain insight into this question, in particular
in connection with the conjectured uniqueness of the Monster theory.

\section*{Acknowledgments}
We thank Karin Baur, J\"urgen Fuchs, Terry Gannon, Hanspeter Kraft,
Marco Matone and Christoph Schweigert for helpful discussions and
correspondences. The research of M.R.G.\ is supported by the Swiss
National Science Foundation; the research of R.V.\ is supported by
an INFN Fellowship.

\appendix

\renewcommand{\theequation}{\Alph{section}.\arabic{equation}}

\section{Decomposition}\label{H2deco}

For the calculation of the trace over $\CFT_2$ of the powers of the quadratic
Casimir $C_2^l$ in section~3.2 it is important to know the decomposition
of $\CFT_2$ with respect to $\g$. If is useful to decompose $\CFT_2$ as
\be
\CFT_2 = \CFT_2^{(0)} \oplus \CFT_2^{{\rm hw}} \ ,
\ee
where $\CFT_2^{(0)}$ are the states at conformal weight two
in the vacuum representation of the affine Lie algebra $\hat\g$, while
$\CFT_2^{\rm hw}$ are the states that are highest weight with respect to
the affine Lie algebra. The states in the vacuum representation can be
determined using the decomposition of the tensor products of the
adjoint. This leads to
\begin{align}
 & \hspace*{-0.2cm} d16\, e8: & \CFT_2^{(0)}  = & 2 \cdot ({\bf
1},{\bf 1})_0 \oplus ({\bf 496},{\bf 1})_{60} \oplus ({\bf 1},{\bf
248})_{60} \oplus ({\bf 496},{\bf 248})_{120} \oplus ({\bf 527},{\bf
1})_{64}\notag   \\*[2pt] & & & \oplus  ({\bf 35960},{\bf 1})_{112}
\oplus
({\bf 1},{\bf 3875})_{96}\notag \\[4pt]
&  \hspace*{-0.2cm}  (e8)^3: & \CFT_2^{(0)} = & 3\cdot ({\bf 1},{\bf
1},{\bf 1})_0 \oplus ({\bf 248},{\bf 1}, {\bf 1})_{60} \oplus ({\bf
3875},{\bf 1},{\bf 1})_{96} \oplus ({\bf 248},{\bf 248},{\bf
1})_{120}\notag \\*[2pt]
& & &  \oplus \hbox{(cycl. perm.)}\notag \\[4pt]
&  \hspace*{-0.2cm}  a17\, e7: & \CFT_2^{(0)} = & 2 \cdot ({\bf
1},{\bf 1})_0 \oplus 2\cdot ({\bf 323},{\bf 1})_{36} \oplus ({\bf
1},{\bf 133})_{36} \oplus ({\bf 323,133})_{72}\notag  \\*[2pt]
& & & \oplus ({\bf 23085},{\bf 1})_{68} \oplus ({\bf 1},{\bf 1539})_{56}\notag \\[4pt]
&  \hspace*{-0.2cm}  d10 \, (e7)^2: & \CFT_2^{(0)}= & 3\cdot ({\bf
1},{\bf 1},{\bf 1})_0 \oplus ({\bf 190},{\bf 1},{\bf 1})_{36} \oplus
({\bf 1},{\bf 133},{\bf 1})_{36} \oplus ({\bf 190},{\bf 133},{\bf
1})_{72}\notag \\*[2pt] & & & \oplus ({\bf 1},{\bf 133},{\bf
133})_{72} \oplus ({\bf 209},{\bf 1},{\bf 1})_{40} \oplus ({\bf
4845},{\bf 1},{\bf 1})_{64} \oplus ({\bf 1},{\bf 1539},{\bf
1})_{56}\notag \\*[2pt]
&&&\oplus \, (2\leftrightarrow 3)\notag
\end{align}
\begin{align}
&  \hspace*{-0.2cm}  a11\, d7\, e6: & \CFT_2^{(0)} = & 3\cdot ({\bf
1},{\bf 1},{\bf 1})_0 \oplus 2 \cdot ({\bf 143},{\bf 1},{\bf
1})_{24} \oplus ({\bf 1},{\bf 91},{\bf 1})_{24} \oplus ({\bf 1},{\bf
1},{\bf 78})_{24}\notag \\*[2pt] & & &  \oplus ({\bf 143},{\bf
91},{\bf 1})_{48}  \oplus ({\bf 143},{\bf 1},{\bf 78})_{48} \oplus
({\bf 1},{\bf 91},{\bf 78})_{48}  \oplus({\bf 4212},{\bf 1},{\bf
1})_{44}\notag  \\*[2pt]
&&& \oplus ({\bf 1},{\bf 104},{\bf 1})_{28}
\oplus ({\bf 1},{\bf 1001},{\bf 1})_{40}
\oplus ({\bf 1},{\bf 1},{\bf 650})_{36}\notag \\[4pt]
&  \hspace*{-0.2cm}  (e6)^4: & \CFT_2^{(0)} = & 4\cdot ({\bf 1},{\bf
1},{\bf 1},{\bf 1})_0 \oplus ({\bf 78},{\bf 1},{\bf 1},{\bf 1})_{24}
\oplus ({\bf 78},{\bf 78},{\bf 1},{\bf 1})_{48} \oplus ({\bf
650},{\bf 1},{\bf 1},{\bf 1})_{36}\notag \\*[2pt]
& & & \oplus\ (\hbox{perm.})\notag \\[4pt]
&  \hspace*{-0.2cm}  (a9)^2\, d6: & \CFT_2^{(0)} = & 3 \cdot ({\bf
1},{\bf 1},{\bf 1})_0 \oplus  2 \cdot ({\bf 99},{\bf 1},{\bf
1})_{20} \oplus ({\bf 1},{\bf 1},{\bf 66})_{20} \oplus ({\bf
99},{\bf 99},{\bf 1})_{40}\notag  \\*[2pt]
& & & \oplus \ ({\bf
99},{\bf 1},{\bf 66})_{40} \oplus ({\bf 1925},{\bf 1},{\bf 1})_{36}
\oplus ({\bf 1},{\bf 1},{\bf 77})_{24} \oplus ({\bf 1},{\bf 1},{\bf
495})_{32}\notag \\*[2pt]
& & & \oplus\ (1\leftrightarrow 2)\notag \\[4pt]
&  \hspace*{-0.2cm}  (d6)^4: & \CFT_2^{(0)} = & 4 \cdot ({\bf
1},{\bf 1},{\bf 1},{\bf 1})_0  \oplus ({\bf 66},{\bf 1},{\bf 1},{\bf
1})_{20} \oplus ({\bf 66},{\bf 66},{\bf 1},{\bf 1})_{40} \oplus
({\bf 77},{\bf 1},{\bf 1},{\bf 1})_{24}\notag \\*[2pt]
& & & \oplus\
({\bf 495},{\bf 1},{\bf 1},{\bf 1})_{32} \oplus
(\hbox{perm.})\notag\\[4pt]
&  \hspace*{-0.2cm} (a5)^4\, d4: & \CFT_2^{(0)} = & 5 \cdot ({\bf
1},{\bf 1},{\bf 1},{\bf 1},{\bf 1})_0  \oplus 2\cdot ({\bf 35},{\bf
1},{\bf 1},{\bf 1},{\bf 1})_{12}  \oplus ({\bf 1},{\bf 1},{\bf
1},{\bf 1},{\bf 28})_{12}\notag  \\*[2pt]
& & & \oplus\ ({\bf
35},{\bf 35},{\bf 1},{\bf 1},{\bf 1})_{24} \oplus ({\bf 35},{\bf
1},{\bf 1},{\bf 1},{\bf 28})_{24}  \oplus ({\bf 189},{\bf 1},{\bf
1},{\bf 1},{\bf 1})_{20}\notag  \\*[2pt]
&&& \oplus\  3\cdot ({\bf 1},{\bf 1},{\bf 1},{\bf 35})_{16}  \oplus (\hbox{perm. $\{1,2,3,4\}$})\notag \\[4pt]
&  \hspace*{-0.2cm}(d4)^6: & \CFT_2^{(0)} = & 6  \cdot ({\bf 1},{\bf
1},{\bf 1},{\bf 1},{\bf 1},{\bf 1})_0  \oplus ({\bf 28},{\bf 1},{\bf
1},{\bf 1},{\bf 1},{\bf 1})_{12}  \oplus ({\bf 28},{\bf 28},{\bf
1},{\bf 1},{\bf 1},{\bf 1})_{24}\notag  \\*[2pt]
&&& \oplus\  3
\cdot ({\bf 35},{\bf 1},{\bf 1},{\bf 1},{\bf 1},{\bf 1})_{16}
\oplus (\hbox{perm.}) \ ,
\end{align}
where the index always denotes the value of the quadratic Casimir.
\smallskip

To determine the contribution from $\CFT_2^{{\rm hw}}$ we recall
that for each simple Lie algebra $\g$, the Sugawara construction
gives \be L_0 = \frac{1}{2(k+h^\vee(\g))} \Bigl( C_2 + 2
\sum_{n=1}^{\infty} J^a_{-n} J^a_n \Bigr) \ , \ee and thus on
highest weight states we have \be C_2 = 2 (k + h^\vee(\g)) L_0 \ .
\ee Since the highest weight states in $\CFT_2^{\rm hw}$ have
conformal dimension $h=2$, it thus follows that \be C_2(\CFT_2^{\rm
hw}) = 4 (k+h^\vee(\g)) \ . \ee For a semi-simple Lie algebra
$\g=\oplus_i \g_i$, this reasoning has to be applied to each factor
separately, but the situation is particularly simple if all $k_i$
and all $h^\vee(\g_i)$ are the same, as is the case for the lattice
theories at $c=24$. For these theories also the dimension of
$\CFT_2^{\rm hw}$ can be easily determined, since one knows that the
total dimension $\dim\CFT_2=196884$, and the dimension of
$\CFT_2^{(0)}$ can be determined as above. This leads to \be
\begin{array}{rll}
d16\, e8:
& \dim\CFT_2^{{\rm hw}}  = 32768 \qquad & C_2(\CFT_2^{\rm hw})=124 \\[2pt]
(e8)^3:
& \dim\CFT_2^{{\rm hw}}  = 0 \qquad &  \\[2pt]
a17\, e7:
& \dim\CFT_2^{{\rm hw}}  = 128520 \qquad & C_2(\CFT_2^{\rm hw})=76 \\[2pt]
d10\, (e7)^2:
& \dim\CFT_2^{{\rm hw}}  = 120064 \qquad & C_2(\CFT_2^{\rm hw})=76 \\[2pt]
a11\,d7\,e6:
& \dim\CFT_2^{{\rm hw}}  = 159194 \qquad & C_2(\CFT_2^{\rm hw})=52 \\[2pt]
(e6)^4:
& \dim\CFT_2^{{\rm hw}}  =  157464 \qquad & C_2(\CFT_2^{\rm hw})=52 \\[2pt]
(a9)^2\,d6:
& \dim\CFT_2^{{\rm hw}}  =  169128 \qquad & C_2(\CFT_2^{\rm hw})=44 \\[2pt]
(d6)^4:
& \dim\CFT_2^{{\rm hw}}  =  168192 \qquad & C_2(\CFT_2^{\rm hw})=44 \\[2pt]
(a5)^4\,d4:
& \dim\CFT_2^{{\rm hw}}  =   184440 \qquad & C_2(\CFT_2^{\rm hw})=28 \\[2pt]
(d4)^6:
& \dim\CFT_2^{{\rm hw}}  =   184320 \qquad & C_2(\CFT_2^{\rm hw})=28 \ .
\end{array}
\ee
With this information it is then straightforward to determine the trace of the powers
of the quadratic Casimir; for example, we have
\be
\begin{array}{llll}
& \hspace*{-0.4cm}d16\,e8:
& \hspace*{-0.2cm} \Tr_{\CFT_2}(C_2^l) = & \hspace*{-0.2cm}
\Bigl[2 \cdot 0^l +
(496+248)\cdot 60^l+(496\cdot 248)\cdot
120^l \\
&&& \quad +\bigl(527\cdot 64^l+ 35960\cdot 112^l\bigr)
+\bigl(3875\cdot 96^l\bigr)\Bigr]+\Bigl[32768\cdot 124^l\Bigr]\\[5pt]
& \hspace*{-0.4cm} (e8)^3:
& \hspace*{-0.2cm}  \Tr_{\CFT_2}(C_2^l) = & \hspace*{-0.2cm}
\Bigl[3\cdot 0^l+ (3\cdot 248)\cdot 60^l+3\cdot(248\cdot 248)\cdot
120^l+3\cdot\bigl(3875\cdot 96^l\bigr)\Bigr] \\[5pt]
& \hspace*{-0.4cm} a17\,e7:
&\hspace*{-0.2cm}   \Tr_{\CFT_2}(C_2^l) = & \hspace*{-0.2cm}
\Bigl[2 \cdot 0^l +
(323+133)\cdot 36^l+(323\cdot 133)\cdot
72^l \\
&&& \quad +\bigl(323\cdot 36^l+ 23085\cdot 68^l\bigr)
+\bigl(1539\cdot 56^l\bigr)\Bigr]+\Bigl[128520\cdot 76^l\Bigr]\\[5pt]
&\hspace*{-0.4cm} d10\,(e7)^2:
& \hspace*{-0.2cm}  \Tr_{\CFT_2}(C_2^l) = & \hspace*{-0.2cm}
\Bigl[3\cdot 0^l + (190+2\cdot133)\cdot 36^l
+(2\cdot190\cdot 133+133^2)\cdot 72^l \\
&&& \quad +\bigl(209\cdot 40^l+4845\cdot 64^l\bigr)
 +2\cdot\bigl(1539\cdot 56^l\bigr)\Bigr]+\Bigl[120064\cdot 76^l\Bigr] \ .
\end{array}
\ee
This then reproduces the results of table
\ref{t:NiemTraces}.

\medskip

Finally, for the Leech lattice theory, the quadratic Casimir is just
the length squared of the underlying lattice vector. At conformal dimension
$h=2$, of the $196884$ states, $324$ are descendants of the vacuum, while
the remaining $196560$ come from the lattice vectors of length squared $4$.
Thus for the Leech theory we simply have
\be
\hbox{Leech:} \qquad
\Tr_{\CFT_2}(C_2^l) = 324\cdot 0^l + 196560 \cdot 4^l \ .
\ee

\section{Riemann surfaces}

\subsection{Riemann period matrices and modular
forms}\label{a:Riemann}

In order to analyse the modular properties of partition functions,
it is useful to define the period matrix of a Riemann surface. Let
$\Sigma$ be a compact Riemann surface of genus $g>0$. Let us define
a basis of the first homology group $H_1(\Sigma,\ZZ)$
$\{\alpha_1,\ldots,\alpha_g,\beta_1,\ldots,\beta_g\}$, with
symplectic intersection matrix
\be
\#(\alpha_i,\alpha_j)=0=\#(\beta_i,\beta_j)\ ,\qquad
\#(\alpha_i,\beta_j)=\delta_{ij}\ ,\qquad i,j=1,\ldots,g\ .
\ee
This condition determines the basis up to a symplectic transformation
\begin{equation}\label{modulhomol}
\Biggl(\begin{matrix}\alpha\\
\beta\end{matrix}\Biggr)\mapsto 
\Biggl(\begin{matrix}\tilde\alpha\\
\tilde\beta\end{matrix}\Biggr):=\Biggl(\begin{matrix}D & C \\ B &
A\end{matrix}\Biggr) \Biggl(\begin{matrix}\alpha\\
\beta\end{matrix}\Biggr)\ ,\qquad\qquad \Biggl(\begin{matrix}A & B
\\ C & D\end{matrix}\Biggr)\in \Sp(2g,\ZZ) \ , \end{equation}
where $\alpha$ und $\beta$ are $g$-dimensional vectors, and
$A,B,C,D$ are $g\times g$ matrices. The choice of such a basis
uniquely determines a basis $\{\omega_1,\ldots,\omega_g\}$ of
holomorphic $1$-differentials normalised with respect to the
$\alpha$-cycles
\be\label{e:modnorm}\oint_{\alpha_i}\omega_j=\delta_{ij}\ ,\qquad
i,j=1,\ldots,g\ . \ee The Riemann period matrix of $\Sigma$ is
then defined by
\be
\Omega_{ij}=\oint_{\beta_i}\omega_j\ ,
\ee
and it has the properties
\be
\Omega_{ij}=\Omega_{ji}\ ,\qquad \im\Omega>0\ .
\ee
Obviously, the basis $\{\omega_1,\ldots,\omega_g\}$, and the Riemann period
matrix depend on the choice of the symplectic basis of
$H_1(\Sigma,\ZZ)$; under the action \eqref{modulhomol} of the
symplectic group, the holomorphic 1-differentials transform as
\begin{subequations}
\begin{align}\label{e:modomega}(\omega_1,\ldots,\omega_g)&
\mapsto(\tilde\omega_1,\ldots,\tilde\omega_g)
=(\omega_1,\ldots,\omega_g)(C\Omega+D)^{-1}\ ,\\
\Omega&\mapsto \tilde\Omega=(A\Omega+B)(C\Omega+D)^{-1}\ .
\end{align}
\end{subequations}
Let us define the Siegel upper half-space as the space of
$g\times g$ symmetric complex matrices with positive
definite imaginary part,
\be
\Sieg_g=\{Z\in M_g(\CC)\mid Z_{ij}=Z_{ji} , \, \im Z >0\}\ .
\ee
The locus $\J_g\subseteq \Sieg_g$ of all
the period matrices of genus $g$ Riemann surfaces is dense in
$\Sieg_g$ for $g\le 3$, whereas for $g>3$ its closure $\bar \J_g$ is
a $(3g-3)$-dimensional subspace of $\Sieg_g$. The quotient
$\J_g/Sp(2g,\ZZ)$ is isomorphic to $\M_g$; in particular, the
Riemann period matrices of two different Riemann surfaces lie in
different $Sp(2g,\ZZ)$-orbits in $\J_g$.

\noindent
A (Siegel) modular form $f$ of degree $g$ and weight $k$ is a holomorphic
function on $\Sieg_g$ such that
\be
f\bigl((AZ+B)(CZ+D)^{-1}\bigr)=\det(CZ+D)^kf(Z)\ ,\quad
M=\Biggl(\begin{matrix}A & B
\\ C & D\end{matrix}\Biggr)\in \Sp(2g,\ZZ)\ .
\ee
For $g=1$, we also require that $f$ is holomorphic at the cusps; a
cusp is a fix-point $p\in\RR\cup\{\infty\}$ under the action of
some $M\in\Sp(2,\ZZ)\cong SL(2,\ZZ)$ with $\Tr(M)=\pm 2$
(a parabolic element). An analogous condition is automatically satisfied for
$g>1$.

\subsection{Degeneration limits and singular Riemann surfaces}\label{s:deglim}

\begin{figure}[htb]\begin{center}
\resizebox{.8\textwidth}{!}{\input{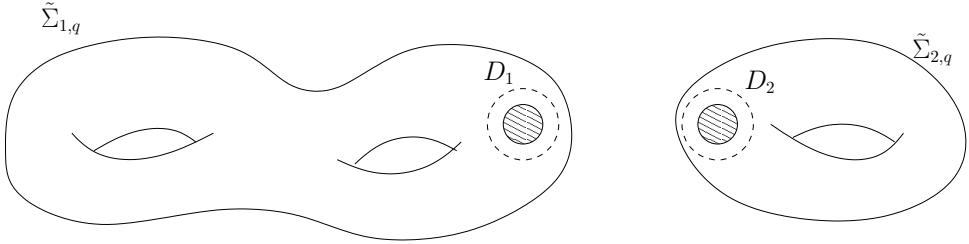}}\end{center}
\caption{\small The surfaces with boundary $\Sigma_{1,q}$ and
$\Sigma_{2,q}$ are the complements of the discs of radius
$|q|^{1/2}$ (the filled discs in the picture) on $\Sigma_1$ and
$\Sigma_2$. The surface $\Sigma_q$ is obtained by sewing together
$\Sigma_{1,q}$ and $\Sigma_{2,q}$, via the identification
\eqref{e:glue} along the boundaries of the discs. The dashed circles
are the boundaries of the coordinate patches $D_1$ and
$D_2$.}\label{f:degcompon}\end{figure}

The moduli space $\M_g$ of smooth Riemann surfaces of genus $g>1$
is the quotient of the Teichm\"uller space, a complex topologically trivial
space of dimension $(3g-3)$, by the discrete mapping class group.
The moduli space $\M_g$ is not compact, and its Deligne-Mumford
compactification $\bar\M_g$ is obtained by adjoining Riemann
surfaces whose only singularities are nodes. In fact, the boundary
$\partial\bar{\M}_g$ is the union of $\lfloor g/2\rfloor+1$ divisors
\be
\partial\bar\M_g=\Delta_0\cup\Delta_1\cup\ldots\Delta_{\lfloor g/2\rfloor }\ ,
\ee where a generic point of $\Delta_k$ corresponds to a Riemann
surface with a node linking two smooth connected components of genus
$k$ and $g-k$, respectively. ($\Delta_0$ is the component where the
node links two points on a single surface of genus $g-1$). In either
case the singular surface is the limit $\lim_{q\to 0}\Sigma_q$ in
$\bar\M_g$, of a suitable family $\{\Sigma_q\}_{0<|q|<1}$ of smooth
Riemann surfaces, parametrised by a complex degeneration parameter
$q\in\CC$. The degenerating surface $\Sigma_q$, $|q|>0$, is defined
by the standard plumbing fixture procedure (see for example
\cite{Fay:1973}), where one identifies (for $k>0$) the boundaries of
local discs  via
\be\label{e:glue}
z_1(p_1)=\frac{q}{z_2(p_2)}\ . \ee Here  $z_i : D_i \rightarrow \CC$
are the local coordinates on some $D_i\subset \Sigma_i$, $i=1,2$,
and \eqref{e:glue} identifies the points $p_i\in D_i$ on the circles
$|z_i(p_i)|=|q|^{1/2}$, $i=1,2$ (see figure \ref{f:degcompon}).
 In the limit $q\to 0$, the
Riemann surface $\Sigma_q$ degenerates to the singular surface
obtained by joining $\Sigma_1$ and $\Sigma_2$, with the points
$u\in\Sigma_1$ and $v\in\Sigma_2$ (that lie at the centres of the
discs $D_1$ and $D_2$, respectively) identified to form a node.

For the case of $\Delta_0$ the only difference is that $u$ and $v$ lie on
the same Riemann surface of genus $g-1$. Similarly, it is clear that we can
also consider a family of smooth curves
$\{\Sigma_{q_1,\ldots,q_n}\}$ depending on $n$ degeneration
parameters $q_i$, $0< |q_i|<1$. As long as the points
$u_1,v_1,\ldots,u_n,v_n$ are pairwise distinct, the limit
$\lim_{q_1,\ldots,q_n\to 0}\Sigma_{q_1,\ldots,q_n}$ is well defined
and corresponds to a singular Riemann surface with $n$ nodes.

\subsection{Schottky uniformisation}\label{a:schottky}

\begin{figure}[htb]\begin{center}
\resizebox{.5\textwidth}{!}{\input{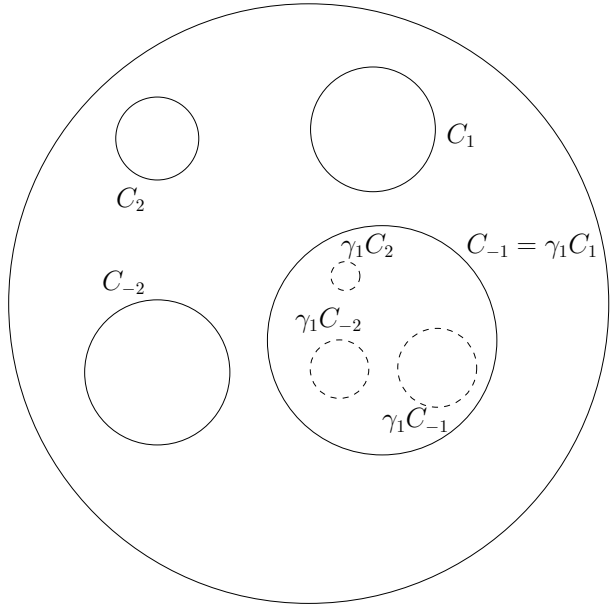}}\end{center}
\caption{\small Schottky uniformization of a Riemann surface of
genus $2$. The fundamental domain $D\subset\hat\CC$ is the
complement of the disks bounded by $C_1,C_{-1},C_2,C_{-2}$. The
Riemann surface is obtained by sewing together $C_1$ with $C_{-1}$
and $C_2$ with $C_{-2}$. The dashed circles are the images of the
cycles $C_{-1},C_{-2},C_2$ under the action of the generator
$\gamma_1$, that maps $C_{1}$ to
$C_{-1}$. The outer circle represents the Riemann sphere
$\hat\CC$.}\label{f:schottky}\end{figure}

A convenient description of genus $g$ Riemann surfaces can be given
in terms of the Schottky uniformisation.  Let $D$ be the open subset of the
Riemann sphere $\hat\CC$, obtained by removing $2g$ closed disks,
with circle boundaries $C_{\pm 1},\ldots,C_{\pm g}$, from $\hat \CC$
(see figure~\ref{f:schottky}). In order to obtain from this a genus $g$ surface, we want
to identify the boundary component $C_r$ with $C_{-r}$, for $r=1,\ldots, g$.
More precisely, let us define $g$ fractional linear transformations
$\gamma_1,\ldots,\gamma_g\in PSL(2,\ZZ)$, such that $\gamma_r$ maps
$C_r$ to $C_{-r}$, for each $r=1,\ldots,g$. We call the discrete subgroup
$\Gamma$ of $PSL(2,\ZZ)$ with distinguished free generators
$\gamma_1,\ldots,\gamma_g$ the marked Schottky group.
It is not difficult to see that $D\in\hat\CC$ is
a fundamental domain for $\Gamma$, and that
$\Sigma$ can be defined as the quotient of the Riemann
sphere by $\Gamma$. (Strictly speaking, we have to exclude the limit points of
fixed points of $\Gamma$.)

All elements of $\Gamma$, and in particular the generators
$\gamma_1,\ldots,\gamma_g$, are loxodromic, {\it i.e.}\ each $\gamma\in\Gamma$ is
conjugate in $PSL(2,\CC)$ to the transformation $z\mapsto q z$
for some multiplier $q$. The multiplier satisfies
$0<|q|<1$ and is uniquely determined by $\gamma$.
More explicitly, we can therefore write $\gamma_r(z)$ as
\be
\frac{\gamma_r(z)-u_r}{\gamma_r(z)-v_r}= q_r\frac{z-u_r}{z-v_r}\ ,
\ee where $0<|q_r|<1$, and $u_r,v_r\in\hat\CC$ are the attracting
and repelling fixed points of $\gamma_r$, respectively. Thus any
marked Schottky group $\Gamma$, and subsequently any Riemann surface
$\Sigma=\Omega/\Gamma$, is completely determined by specifying the
multipliers and the attracting and repelling points of its
generators. For $g>1$, we can apply an overall $PSL(2,\CC)$
conjugation to fix $u_g=0$, $v_g=\infty$, $u_{g-1}=1$; the resulting
Schottky group is called normalised. The space of normalised marked
Schottky groups defines the Schottky space $\Sch$. It is a
$(3g-3)$-dimensional complex manifold parameterised by
\be
q_1,\ldots,q_g,u_1,\ldots,u_{g-2},v_1,\ldots,v_{g-1}\ , \ee and it
defines a finite covering of the moduli space of Riemann surfaces.
The curves $C_1,\ldots,C_g$ can be taken to define the cycles
$\alpha_1,\ldots,\alpha_g$ in a symplectic basis of
$H_1(\Sigma,\ZZ)$ (see appendix \ref{a:Riemann}). It follows that
the choice of a Schottky group uniformising a Riemann surface
$\Sigma$ canonically determines a basis
$\{\omega_1,\ldots,\omega_g\}$ of holomorphic $1$-differentials on
$\Sigma$, satisfying the normalisation condition \eqref{e:modnorm}.
\smallskip

For $g=1$, the Schottky group is a discrete abelian subgroup
$\Gamma\cong \ZZ$ of $PSL(2,\CC)$, freely generated by a loxodromic
element $\gamma$. By a $PSL(2,\CC)$-conjugation the attracting and
repelling points of $\gamma$ can be fixed to $0$ and $\infty$
respectively, so that $\gamma:z\mapsto qz$, for some $q\in\CC$,
$0<|q|<1$. The modular parameter $\tau$ is related to $q$ by
$q=e^{2\pi i\tau}$; the coordinate $w$ on the usual torus
$w\in\CC/(\ZZ+\tau\ZZ)$ is related to the coordinate $z$ by
\begin{equation}\label{e:flattoSchottky}
z(w)=q^{1/2}e^{2\pi i w}\ ,
\end{equation}
so that
\be
z(w+1)=z(w)\ ,\qquad z(w+\tau)=\gamma(z(w))\ .
\ee

Finally, a family $\Sigma_q$ of Riemann surfaces of genus $g$
degenerating, in the limit $q\to 0$, to a singular surface in
$\Delta_0$, can be easily described in terms of the Schottky
uniformisation. Let us define a Schottky group $\Gamma_q$ with
generators $\gamma_1,\ldots,\gamma_{g-1},\gamma_g(q)$ and such that
the multiplier $q_g$ of $\gamma_g$ equals the degeneration parameter
$q_g=q$. The limit $q\to 0$ corresponds then to pinching the
homologically non-trivial cycle $C_g$ down to a point.

\section{Evidence for the Lie algebra conjecture}
\label{s:conj}

Recall the conjecture of section~5.3: if  $R_1$ and $R_2$ are two irreducible
representations of a simple Lie algebra $\g$ such that
the eigenvalues of $C_l^{(\g)}$ on $R_i$ are equal,
\be\label{condl1}
C_l^{(g)} (R_1) =  C_l^{(g)} (R_2) \qquad
\hbox{for all $l$,}
\ee
where $C_l^{(\g)}$ is the Casimir operator defined in (\ref{e:AdCas}),
then either $R_1\cong R_2$ or $R_1\cong \pi(R_2)$, where $\pi$ is a
non-trivial (outer)  automorphism of $\g$.
\smallskip

Let us collect some support for this conjecture. The situation is obviously simplest if the
algebra generated by the $C_l^{(\g)}$ is equivalent to the algebra generated by the
$C_l^\perp$. Then the usual analysis for the invariant algebra shows that
(\ref{condl1}) implies $R_1\cong R_2$.

The two algebras are the same if
all $I_l(\g)\neq 0$ and if the different Casimir operators $C^\perp_l$ have
different degrees. Indeed then we can use (\ref{Crel}) recursively
to express the generators $C_l^\perp$ in terms of $C_l^{(\g)}$, thus establishing
that the algebra generated by the $C_l^\perp$ is a subalgebra of the algebra
generated by the $C_l^{(\g)}$, and hence isomorphic to it.
The above condition is satisfied for the simple Lie
algebras  $b(r)$, $c(r)$, $e7$, $e8$, $f4$ and $g2$. All of them do not have any
non-trivial outer automorphisms, and thus $R_1\cong R_2$ is the only possibility.
The other cases are more difficult, so let us deal with them in turn.

\subsection{$d(r)$ algebras}

For the $d(r)$ algebras, the independent Casimirs have degrees
$2,4,6,\ldots,2r-2,r$. The analysis depends a bit on whether $r$ is
even or odd.
\smallskip

\noindent {\bf $r$ odd:} If $r$ is odd, then all the Casimir operators $C_l^\perp$
have different degree, but
for the Casimir of odd degree $r$ the index $I_r(\g)$ vanishes. In fact, the index
always vanishes for Casimir operators of odd degree since the generators of the
adjoint representations are anti-symmetric, $t^a = - (t^a)^T$, and thus
\begin{eqnarray}
\Tr_{ad}(t^{a_l}\cdots t^{a_1}) & = &
(-1)^l\Tr_{ad}((t^{a_l})^T\cdots (t^{a_1})^T)
=(-1)^l\Tr_{ad}((t^{a_1}\cdots t^{a_l})^T) \nonumber \\
& = & (-1)^l\Tr_{ad}(t^{a_1}\cdots t^{a_l})\ .
\end{eqnarray}
Since $c^{a_1\ldots a_l}$ is totally symmetric, it then follows that
\be\label{lodd}
c^{a_1\ldots a_l}\Tr_{ad}(t^{a_1}\cdots t^{a_l})
=c^{a_1\ldots a_l}\Tr_{ad}(t^{a_l}\cdots t^{a_1})
=(-1)^lc^{a_1\ldots a_l}\Tr_{ad}(t^{a_1}\cdots t^{a_l})\ ,
\ee
thus showing that the index $I_l(\g)$ vanishes if $l$ is odd.

For the case of $d(r)$ one can show by an
explicit calculation that the algebra generated by the $C_l^{(\g)}$ coincides with
the subalgebra of the invariant algebra generated by
\be
C^\perp_2,\ \ldots,\ C^\perp_{2r-2},\ (C_r^\perp)^2\ .
\ee
This allows us to distinguish all representations, except those that differ
by the sign of the eigenvalue of $C_r^\perp$. One can show that two
representations that only differ by the sign of the eigenvalue of $C_r^\perp$
are precisely charge conjugate representations. Thus we can identify
representations up to charge conjugation, in agreement with the conjecture.
\medskip

\noindent {\bf $r$ even:} For $r$ even, all the Casimir
operators have even degree, but there are now two independent
Casimirs of degree $r$, which we denote by $C_r^\perp$ and
$\tilde C_r^\perp$. We choose the convention that the invariant
$\tilde C_r^\perp$ of degree $r$ is only non-zero for the spinor
representations, {\it i.e.}\ the representations that are not
representations of $SO(2r)$. It then follows that
$\tilde{I}_r(\g)=0$, whereas it can be shown that the index
$I_l(\g)$, $l=2,\ldots,2r-2$, is related to the analogous index
$I_l(V)$ for the vector representation by
\be
I_l(\g)=(2r-2^{l-1})I_l(V)\ .
\ee
It is known that $I_l(V)\neq 0$ for all
$l=2,\ldots,2r-2$, so that, if we restrict to the case where $r$ is
not an even power of $2$, we obtain $I_l(\g)\neq 0$ as well.
Provided that $r\neq 4^n$ one can then show that the algebra
generated by the $C_l(\g)$ coincides with the subalgebra of the
invariant algebra generated by
\be
C^\perp_2,\ \ldots,\ C^\perp_{2r-2},\ (\tilde{C}_r^\perp)^2\ .
\ee
This allows one to distinguish all representations, except
those that differ by the sign of the eigenvalue of
$\tilde{C}_r^\perp$, {\it i.e.}\ up to the outer automorphism
corresponding to spin flip.

The case $r=4^n$ includes in particular $d4=so(8)$, where we know
that something special has to happen (since this algebra has an
enhanced triality symmetry). In fact, for
$d4$, both the fourth order indices $I_4(\g)$ and
$\tilde I_4(\g)$ vanish. Unfortunately, we have not been
able to show that for $r=4^n$ with $r\neq 4$, the algebra
generated by $C_l^\g$ is sufficient to distinguish irreducible
representations up to spin flip. (However, we are also not aware
of any counterexample.)

\subsection{$e6$ algebra}

For the $e6$ algebra, the degrees of the  independent Casimirs are
$2,5,6,8,9,12$. The indices $I_l(\g)$ are non-zero for all the even
$l$. One can show that the subalgebra generated by the $C_l^{\g}$
is precisely the subalgebra of the full invariant algebra generated by
\be
C^\perp_2,\ C^\perp_6,\ C^\perp_8,\ (C^\perp_5)^2,\ C^\perp_{12},\
C^\perp_5C^\perp_9,\ (C^\perp_9)^2 \ . \ee
This allows one to identify all representations up to charge conjugation.

\subsection{$a(r)$ algebras}

The case of the $a(r)$ algebras is the most complicated, because there
are several Casimirs of odd degree. More precisely, the independent
Casimirs have degree $2,3,4,\ldots,r+1$; the index $I_l(\g)$ of all the
Casimirs of even degree is non-zero, but because of (\ref{lodd})
$I_l(\g)=0$ for all odd $l$. In analogy with the $d(r)$ and
$e6$ cases, it is natural to expect that the subalgebra generated by the
$C_l^{(\g)}$ contains
\be
C_2^\perp,\ C^\perp_4,\ C^\perp_6,\ \ldots,\
C^\perp_{2\lfloor(r+1)/2\rfloor},\ (C^\perp_3)^2, \
C^\perp_3C^\perp_5,\ \ldots,\ C^\perp_3C^\perp_{2\lfloor
r/2\rfloor+1}\ . \ee This can be proved for $r\le 4$, but we have
not managed to establish it in general. If true, it would imply that
we can identify representations up to charge conjugation.

\end{document}